\documentclass[pdflatex,sn-mathphys]{sn-jnl}% Math and Physical Sciences 
\jyear{2021}
\theoremstyle{thmstyleone}%
\theoremstyle{thmstyletwo}%
\theoremstyle{thmstylethree}%
\usepackage{microtype}
\begin{document}
\title[Recent Advances and Applications of Deep Learning Methods in Materials Science]{Recent Advances and Applications of Deep Learning Methods in Materials Science}
\author*[1,2]{\fnm{Kamal} \sur{Choudhary}}\email{kamal.choudhary@nist.gov}
\author[3]{\fnm{Brian} \sur{DeCost}}
\author[4]{\fnm{Chi} \sur{Chen}}
\author[5]{\fnm{Anubhav} \sur{Jain}}
\author[1]{\fnm{Francesca} \sur{Tavazza}}
\author[6]{\fnm{Ryan} \sur{Cohn}}
\author[7]{\fnm{Cheol} \sur{WooPark}}
\author[8]{\fnm{Alok} \sur{Choudhary}}
\author[8]{\fnm{Ankit} \sur{Agrawal}}
\author[9]{\fnm{Simon J. L.} \sur{Billinge}}
\author[6]{\fnm{Elizabeth} \sur{Holm}}
\author[4]{\fnm{Shyue Ping} \sur{Ong}}
\author[7]{\fnm{Chris} \sur{Wolverton}}

\affil[1]{\orgdiv{Materials Science and Engineering Division}, \orgname{National Institute of Standards and Technology}, \orgaddress{\city{Gaithersburg}, \postcode{20899}, \state{MD}, \country{USA}}}

\affil[2]{ \orgname{Theiss Research}, \orgaddress{\city{La Jolla}, \postcode{92037}, \state{CA}, \country{USA}}}

\affil[3]{\orgdiv{Material Measurement Science Division}, \orgname{National Institute of Standards and Technology}, \orgaddress{\city{Gaithersburg}, \postcode{20899}, \state{MD}, \country{USA}}}

\affil[4]{\orgdiv{Department of NanoEngineering}, \orgname{University of California San Diego}, \orgaddress{ \postcode{92093}, \state{CA}, \country{USA}}}

\affil[5]{\orgdiv{Energy Technologies Area, Lawrence Berkeley National Laboratory, Berkeley, CA, USA}}

\affil[6]{\orgdiv{Department of Materials Science and Engineering, Carnegie Mellon University, Pittsburgh, PA, 15213, USA}}

\affil[7]{\orgdiv{Department of Materials Science and Engineering, Northwestern University, Evanston, IL, 60208, USA}}

\affil[8]{\orgdiv{Department of Electrical and Computer Engineering, Northwestern University, Evanston, IL, 60208, USA}}

\affil[9]{\orgdiv{Department of Applied Physics and Applied Mathematics and the Data Science Institute, Fu Foundation School of Engineering and Applied Sciences, Columbia University, New York, NY, 10027, USA}}

\abstract{Deep learning (DL) is one of the fastest growing topics in materials data science, with rapidly emerging applications spanning atomistic, image-based, spectral, and textual data modalities. DL allows analysis of unstructured data and automated identification of features. Recent development of large materials databases has fueled the application of DL methods in atomistic prediction in particular. In contrast, advances in image and spectral data have largely leveraged synthetic data enabled by high quality forward models as well as by generative unsupervised DL methods. In this article, we present a high-level overview of deep-learning methods followed by a detailed discussion of recent developments of deep learning in atomistic simulation, materials imaging, spectral analysis, and natural language processing. For each modality we discuss applications involving both theoretical and experimental data, typical modeling approaches with their strengths and limitations, and relevant publicly available software and datasets. We conclude the review with a discussion of recent cross-cutting work related to uncertainty quantification in this field and a brief perspective on limitations, challenges, and potential growth areas for DL methods in materials science. The application of DL methods in materials science presents an exciting avenue for future materials discovery and design.}

\maketitle

\section{Introduction}\label{sec:intro}

``Processing-structure-property-performance" is the key mantra in Materials Science and Engineering (MSE) \cite{callister2021materials}. The length and time scales of material structures and phenomena vary significantly among these four elements, adding further complexity \cite{saito2013computational}. For instance, structural information can range from detailed knowledge of atomic coordinates of elements to the microscale spatial distribution of phases (microstructure), to fragment connectivity (mesoscale), to images and spectra. Establishing linkages between the above components is a challenging task.

Both experimental and computational techniques are useful to identify such relationships. Due to rapid growth in automation in experimental equipments and immense expansion of computational resources, the size of public materials datasets has seen an exponential growth. Through the Materials Genome Initiative (MGI) \cite{de2019new} and the increasing adoption of Findable, Accessible, Interoperable, Reusable (FAIR) \cite{wilkinson2016fair} principles, several large experimental and computational datasets have been developed \cite{choudhary2020joint,kirklin2015open,jain2013commentary,curtarolo2012aflow,ramakrishnan2014quantum,draxl2018nomad,wang2005pdbbind,zakutayev2018open}. Such an outburst of data requires automated analysis which can be facilitated by machine-learning (ML) techniques \cite{friedman2001elements,agrawal2016perspective,vasudevan2019materials,schmidt2019recent,butler2018machine,xu2020deep,schleder2019dft,agrawal2019deep}. 

Deep learning (DL) \cite{Goodfellow-et-al-2016,lecun2015deep} is a specialized branch of machine learning (ML). Originally inspired by biological models of computation and cognition in the human brain~\cite{McCulloch,rosenblatt}, one of DL's major strengths is its potential to extract higher-level features from the raw input data.

DL applications are rapidly replacing conventional systems in many aspects of our daily lives as, for example, in image and speech recognition, web search, fraud detection, email/spam filtering, financial risk modeling, and so on. DL techniques have been proven to provide exciting new capabilities in numerous fields (such as playing Go \cite{gibney2016google}, self-driving cars \cite{ramos2017detecting}, navigation, chip design, particle physics, protein science, drug discovery, astrophysics, object recognition \cite{buduma2017fundamentals}, etc). 

Recently DL methods have been outperforming other machine learning techniques in numerous scientific fields, such as chemistry, physics, biology, and materials science \cite{kearnes2016molecular,albrecht2017deep,ge2020deep,agrawal2019deep,agrawal2020materials,erdmann2021deep}. DL applications in MSE are still relatively new, and the field has not fully explored its potential, implications, and limitations. DL provides new approaches for investigating material phenomena and has pushed materials scientists to expand their traditional toolset.

DL methods have been shown to act as a complementary approach to physics based methods for materials design. While large datasets are often viewed as a prerequisite for successful DL applications, techniques such as transfer learning, multi-fidelity modelling, and active learning can often make DL feasible for small datasets as well \cite{chen2019graph,jha2019enhancing,cubuk2019screening,chenLearningPropertiesOrdered2021}. 

Traditionally, materials have been designed experimentally using trial and error methods with a strong dose of chemical intuition. In addition to being a very costly and time consuming approach, the number of material combinations is so huge that it is intractable to study experimentally, leading to the need for empirical formulation and computational approaches. While computational approaches (such as density functional theory, molecular dynamics, Monte Carlo, phase-field,  finite elements) are much faster and cheaper than experiments, they are still limited by length and time scale constraints, which in turn limits their respective domains of applicability. DL methods can offer substantial speedups compared to conventional scientific computing, and, for some applications, are reaching an accuracy level comparable to physics-based or computational models.

Moreover, entering a new domain of materials science and performing cutting-edge research requires years of education, training, and development of specialized skills and intuition. Fortunately, we now live in an era of increasingly open data and computational resources. Mature, well-documented DL libraries makes DL research much more easily accessible to newcomers than almost any other research field. Testing and benchmarking methodologies such as underfitting/overfitting/cross-validation \cite{vasudevan2019materials,schmidt2019recent,artrith2021best} are common knowledge, and standards for measuring model performance are well established in the community.
 
Despite their many advantages, DL methods have disadvantages too, the most significant one being their black-box nature \cite{holm2019defense} which may hinder physical insights into the phenomena under examination. Evaluating and increasing  interpretability and explainability of DL models still remains an active field of research. Generally a DL model has a few thousands to millions of parameters, making model interpretation and direct generation of scientific insight difficult.

Although there are several good recent reviews of ML applications in MSE \cite{vasudevan2019materials,schleder2019dft,schmidt2019recent,mueller2016machine,wei2019machine,butler2018machine,liu2020machine,wang2020machine,morgan2020opportunities,himanen2019data,rajan2013informatics,montans2019data,aykol2019materials,stanev2021artificial,chen2020critical}, DL for materials has been advancing rapidly, warranting a dedicated review to cover the explosion of research in this field. In this article, we discuss some of the basic principles in DL methods and then highlight major trends among the recent advances in DL applications for materials science. As the tools and datasets for DL applications in materials keep evolving, we provide a github repository (\url{https://github.com/deepmaterials/dlmatreview}) that can be updated as new resources are made publicly available.

\section{Basics of deep learning}\label{sec:basics}

\subsection{General machine learning concepts}\label{sec:general-concepts}

Artificial intelligence (AI) \cite{friedman2001elements} is the development of machines and algorithms that mimics human intelligence, for example, by optimizing actions to achieve certain goals. Machine learning (ML) is a subset of AI, and provides the ability to learn without explicitly being programmed for a given dataset such as playing chess, social network recommendation etc. DL, in turn, is the subset of ML that takes inspiration from biological brains and uses multi-layer neural networks to solve ML tasks. A schematic of AI-ML-DL context and some of the key application areas of DL in materials science and engineering field are shown in Fig. 1.

Some of the commonly used ML technologies are linear regression, decision trees and random forest in which generalized models are trained to learn coefficients/weights/parameters for a given dataset (usually structured i.e., on a grid or a spreadsheet). 

For unstructured data  (such as pixels or features from an image, sounds, text and graphs) applying traditional ML techniques becomes challenging because users have to first extract generalized meaningful representations or features themselves (such as calculating pair-distribution for an atomic structure) and then train the ML models. Hence, the process becomes time consuming, brittle and not easily-scalable. Here, deep learning (DL) techniques become more important. 
 
 DL methods are based on artificial neural networks and allied techniques. According to the ``universal approximation theorem" \cite{cybenko1989approximation,kidger2020universal}, neural networks can approximate any function to arbitrary accuracy.

\begin{figure}
    \centering
    \includegraphics[trim={0 0cm 0 0cm},clip,width=1.0\textwidth]{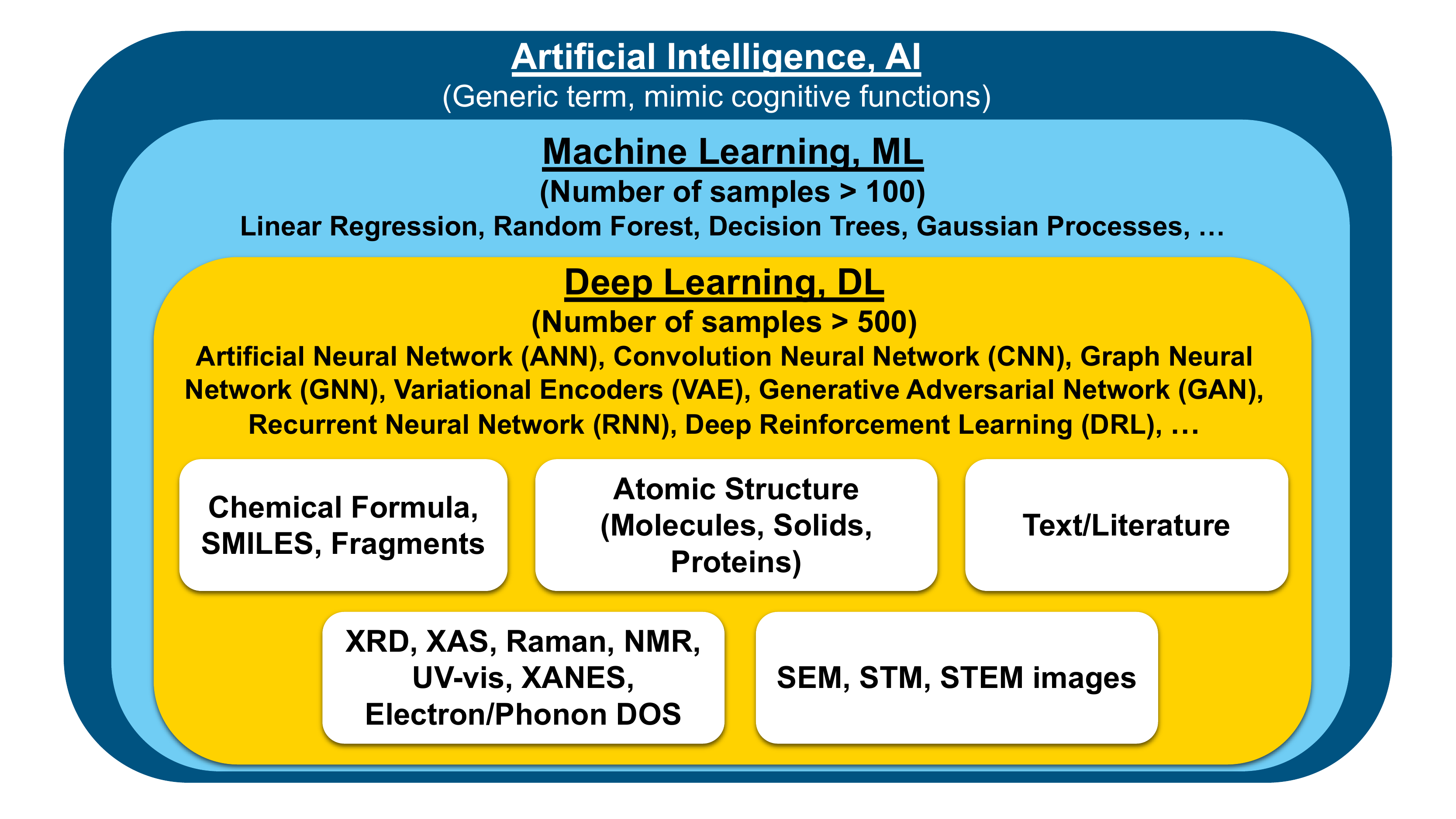}
    \caption{Schematic showing an overview of Artificial Intelligence (AI), Machine Learning (ML) and Deep Learning (DL) methods and its applications in materials science and engineering. Deep learning is considered as a part of machine-learning which is contained in an umbrella term artificial intelligence.}
\end{figure}

\subsection{Neural networks}\label{sec:neural-nets}
\subsubsection{Perceptron}

A perceptron or a single artificial neuron \cite{minsky2017perceptrons} is the building block of artificial neural networks (ANNs) and performs forward propagation of information. For a set of inputs $[x{_1},x{_2},...,x{_m}]$ to the perceptron, we assign floating number weights (and biases to shift wights) $[w{_1},w{_2},...,w{_m}]$ and then we multiply them correspondingly together to get a sum of all of them. Some of the common software packages \cite{nistdisclaimer} allowing NN trainings are: PyTorch \cite{paszke2019pytorch}, Tensorflow \cite{abadi2016tensorflow} and MXNet \cite{chen2015mxnet}.

\subsubsection{Activation function}

 Activation functions (such as sigmoid, hyperbolic tangent (tanh), rectified linear unit (ReLU), leaky ReLU, Swish) are the critical nonlinear components that enable neural networks to compose many small building blocks to learn complex nonlinear functions. For example, the sigmoid activation maps real numbers to the range (0, 1); this activation function is often used in the last layer of binary classifiers to model probabilities. The choice of activation function can affect training efficiency as well as final accuracy~\cite{DBLP:journals/corr/abs-1811-03378}.

\subsubsection{Loss function, gradient descent and normalization}
The weight matrices of a neural network are initialized randomly or obtained from a pre-trained model. These weight matrices are multiplied with the input matrix (or output from a previous layer) and subjected to a nonlinear activation function to yield updated representations, which are often referred to as activations or feature maps.
The loss function (also known as objective function or empirical risk) is calculated by comparing the output of the neural network and the known target value data.
Typically, network weights are iteratively updated via stochastic gradient descent algorithms to minimize the loss function until desired accuracy is achieved.
Most modern deep learning frameworks facilitate this by using reverse-mode automatic differentiation~\cite{JMLR:v18:17-468} to obtain the partial derivatives of loss function with respect to each network parameter through recursive application of the chain rule.
Colloquially, this is also known as back-propagation. 

Some of the common gradient descent algorithms are: Stochastic Gradient Descent (SGD), Adam, Adagrad etc. The learning rate is an important parameter in gradient descent. Except SGD, all other methods use adaptive learning parameter tuning. Depending on the objective such as classification or regression, different loss functions such as Binary Cross Entropy (BCE), Negative Log likelihood (NLLL) or Mean Squared Error (MSE) are used.

The inputs of a neural network are generally scaled i.e., normalized to have zero mean and unit standard deviation. Scaling is also applied to the input of hidden layers (using batch or layer normalization) to improve the stability of ANNs. 

\subsubsection{Epoch and mini batches}
A single pass of the entire training data is called an epoch, and multiple epochs are performed until the weights converge. In DL, datasets are usually large and computing gradients for the entire dataset and network becomes challenging. Hence, the forward passes are done with small subsets of the training data called mini-batches. 

\subsubsection{Underfitting, overfitting, regularization and early stopping}
During an ML training, the dataset is split into training, validation and test sets. The test set is never used during the training process. A model is said to be underfitting if the model performs poorly on training set and lacks capacity to fully learn the training data. A model is said to overfit if the model performs too well on the training data but does not perform well on the validation data. Overfitting is controlled with regularization techniques such as dropout and early stopping. 

Regularization discourages the model from simply memorizing the training data so that the model can be generalizable. One of the most popular regularizations is dropout in which we randomly set the activations for an NN layer to zero. 

In early stopping, further epochs for training are stopped before the model overfits i.e., accuracy on the validation set flattens or decreases. 

\subsection{Convolution neural networks}\label{sec:cnn}
Convolutional neural networks (CNN) \cite{lecun1995convolutional}  can be viewed as a regularized version of multilayer perceptrons with a strong inductive bias for learning translation-invariant image representations. There are four main components in CNNs: a) learnable convolution filterbanks, b) nonlinear activations, c) spatial coarsening (via pooling or strided convolution), d) a prediction module, often consisting of fully-connected layers that operate on a global instance representation. 

In CNNs we use convolution functions with multiple kernels or filters with trainable and shared weights or parameters, instead of general matrix multiplication. These filters/kernels are matrices with a relatively small number of rows and columns that convolve over the input to automatically extract high-level local features in the form of feature maps. The filters slide/convolve (element wise multiply) across the input with a fixed number of strides to produce the feature map and the information thus learnt is passed to the hidden/fully-connected layers. These filters can be one, or two or three dimensional depending on the input data. 

Similar to the fully connected NNs, nonlinearities such as ReLU are then applied that allows us to deal with non-linear and complex data. The pooling operation preserves spatial invariance, downsamples and reduces dimension of each feature map obtained after convolution. These downsampling/pooling operations can be of different types such as maximum-pooling, minimum-pooling, average pooling and sum pooling. 
After one or more convolutional and pooling layers, the outputs are usually reduced to a one-dimensional global representation.
CNNs are especially popular for image data.

\subsection{Graph neural networks}\label{sec:gnn}

\subsubsection{Graphs and their variants}
Classical CNNs as described above are based on a regular grid Euclidean data (such as 2D grid in images). However, real life data-structures such as social networks, segments of images, word-vectors, recommender systems and atomic/molecular structures are usually non-Euclidean. In such cases, graph based non-Euclidean data-structures become specially important.

Mathematically, a graph G is defined as a set of nodes/vertices V, a set of edges/links, E and node features, X: $G = (V,E,X)$ \cite{wilson1979introduction,west2001introduction,wang2019deep} and can be used to represent non-Euclidean data. An edge is formed between a pair of two nodes and contains the relation information between the nodes. Each node and edge can have attributes/features associated with it. An adjacency matrix A is a square matrix  indicating if there are connections between the nodes or not in the form of 1 and 0. A graph can be of various types such as: undirected/directed, weighted/unweighted, homogeneous/heterogeneous, static/dynamic. 

An undirected graph captures symmetric relations between nodes, while an directed one captures asymmetric relations such that $Aij \neq Aji$. In a weighted graph, each edge is associated with a scalar weight rather than just 1s and 0s. In a homogeneous graph, all the nodes represent instances of the same type and all the edges capture relations of the same type while in a heterogeneous graph, the nodes and edges can be of different types. Heterogeneous graphs provide an easy interface for managing nodes and edges of different types as well as their associated features. When input features or graph topology vary with time, they are called dynamic graphs otherwise they are considered static. If a node is connected to another node more than once it is termed as a multi-graph.
%https://arxiv.org/ftp/arxiv/papers/1812/1812.08434.pdf
%https://docs.dgl.ai/en/0.6.x/tutorials/models/index.html

\subsubsection{Types of GNNs}
At present, GNNs are probably the most popular AI method for predicting various materials properties based on structural information.
Graph neural networks (GNNs) are DL methods that operate on graph domain and can capture the dependence of graphs via message passing between the nodes and edges of graphs. There are two key steps in GNN training: a) we first aggregate information from neighbors and b) update the nodes and/or edges. Importantly, aggregation is permutation invariant. Similar to the fully-connected NNs, the input node-features, X (with embedding matrix) are multiplied with the adjacency matrix and the weight matrices and then multiplied with the non-linear activation function to to provide outputs for the next layer. This method is called propagation rule.

Based on the propagation rule and aggregation methodology, there could be different variants of GNNs such as Graph convolutional network (GCN)~\cite{kipf2016semi}, Graph attention network (GAT)~\cite{velivckovic2017graph}, Relational-GCN~\cite{schlichtkrull2017modeling}, graph recurrent network (GRN)~\cite{song-etal-2018-graph}, Graph isomerism network (GIN)~\cite{xu2018powerful}, and Line graph neural network (LGNN)~\cite{chen2017supervised}.
Graph convolutional neural networks are the most popular GNNs. 

\subsection{Sequence-to-sequence models}\label{sec:sequence}

Traditionally, learning from sequential inputs such as text involves first generating a fixed-length input from the data. For example, the ``bag-of-words" approach simply counts the number of instances of each word in a document and produces a fixed-length vector that is the size of the overall vocabulary.

In contrast, sequence-to-sequence models can take into account sequential / contextual information about each word and produce outputs of arbitrary length. For example, in named entity recognition (NER), an input sequence of words (e.g., a chemical abstract) is mapped to an output sequence of ``entities" or categories where every word in the sequence is assigned a category. 

An early form of sequence-to-sequence model is the recurrent neural network, or RNN. Unlike the fully connected NN architecture, where there is no connection between hidden nodes in the same layer, but only between nodes in adjacent layers, RNN have feedback connections and each hidden layer can be unfolded and processed similarly to traditional NNs sharing same weight matrices. There are multiple types of RNNs, of which the most common ones are: gated recurrent unit recurrent neural network (GRURNN), long short-term memory (LSTM) network, and clockwork RNN (CW-RNN) \cite{jing2018deep}. 

However, all such RNNs suffer from some drawbacks, including: (i) difficulty of parallelization and therefore difficulty in training on large data sets and (ii) difficulty in preserving long-range contextual information due to the ``vanishing gradient" problem. Nevertheless, as we will later describe, LSTMs have been successfully applied to various NER problems in the materials domain.

More recently, sequence-to-sequence models based on a ``transformer" architecture, such as Google's Bidirectional Encoder Representations from Transformers (BERT) model \cite{devlin2018bert,nistdisclaimer}, have helped address some of the issues of traditional RNNs. Rather than passing a state vector that is iterated word-by-word, such models use an attention mechanism to allow access to all previous words simultaneously without explicit time steps. This facilitates parallelization and also better preserves long-term context. 

\subsection{Deep generative models (VAE and GAN)}\label{sec:generative}

While the above DL frameworks are based on supervised machine learning (i.e., we know the target or ground truth data such as in classification and regression) and discriminative (i.e., learn differentiating features between various datasets), many AI tasks are based on unsupervised (such as clustering) and are generative (i.e., aim to learn underlying distributions). 

Generative models are used to a) generate data samples similar to the training set with variations i.e., augmentation, b) learn good generalized latent features, c) guide mixed reality applications such as virtual try-on. There are various types of generative models, of which the most common are: a) variational encoders (VAE), which explicitly define and learn likelihood of data, b) Generative adversarial networks (GAN), which learn to directly generate samples from model's distribution, without defining any density function.

\subsection{Deep reinforcement learning}\label{sec:rl}

Reinforcement learning (RL) deals with tasks in which a computational agent learns to make decisions by trial and error. Deep RL uses DL into the RL framework, allowing agents to make decisions from unstructured input data. In traditional RL, Markov decision process (MDP) is used in which an agent at every timestep takes action to receive a scalar reward and transitions to the next state according to system dynamics to learn policy in order to maximize returns. However, in deep RL, the states are high-dimensional (such as continuous images or spectra) which act as an input to DL methods. DRL architectures can be either model based or model free.

\section{Applications of DL methods}\label{sec:applications}
Some aspects of successful DL application that require materials-science-specific considerations are:
1) acquiring large, balanced and diverse datasets (often on the order of 10000 data points or more),
2) determing an appropriate DL approach and suitable vector or graph representation of the input samples, and
3) selecting appropriate performance metrics relevant to scientific goals.
%  such as MAE \& MSE  (regression)/ F1 score (classification) / reconstruction loss (VAE) etc

% brian: IMO these things are important for executing DL research well, but are not 
% specific to materials science research
% 4) tuning the hyper-parameters for the DL model, 
% 5) splitting the entire dataset in to train-test or train-validation-test (preferred) splits as generally new data won't be generated to test the model,
% 6) monitor learning curves to avoid overfitting.

In the following sections we discuss some of the key areas of materials science in which DL has been applied with available links to repositories and datasets that help in reproducibility and extensibility of the work.
In this review we categorize materials science applications at a high level by the type of input data considered: \ref{sec:atomistic} atomistic, \ref{sec:stoichiometric} stoichiometric, \ref{sec:spectral} spectral, \ref{sec:image} image, and \ref{sec:nlp} text.
Within each broad materials data modality, we summarize prevailing machine learning tasks and their impact on materials research and development.

%% goal? summarize big conceptual advances since the latest review

\subsection{Atomistic and chemical representations}\label{sec:atomistic}
 In this section we provide a few examples of solving materials science problems with DL methods trained on atomistic data. Atomic structure of a material usually consists of atomic coordinates and atomic composition information of a material. Arbitrary number of atoms and types of elements in a system poses a challenge to apply traditional ML algorithms for atomistic predictions. DL based methods are an obvious strategy to tackle this problem. There have been several previous attempts to represent crystals and molecules using fixed size descriptors such as Coulomb matrix \cite{rupp2012fast,bartok2013representing,faber2017prediction}, classical force field inspired descriptors (CFID) \cite{choudhary2018machine,choudhary2021high,choudhary2020data}, pair distribution function (PRDF), Voronoi tessellation \cite{ward2017including,isayev2017universal,liu2019using}. Recently graph neural network methods have been shown to surpass previous hand-crafted feature set \cite{kearnes2016molecular}. 
 
 DL for atomistic materials applications include: a) force-field development, b) direct property predictions, c) materials screening. In addition to the above points, we also elucidate upon some of the recent generative adversarial network and complimentary methods to atomistic aproaches.

 \subsubsection{Databases and software libraries}
In Table~\ref{tab:atomistic-deep-learning} we provide some of the commonly used datasets used for atomistic DL models for molecules, solids and proteins. We note that the computational methods method used for different datasets are different and many of them are continuously evolving. Generally it takes years to generate such databases using conventional methods such as density functional theory, while DL methods can be used to make predictions with much reduced computational cost and reasonable accuracy.

\begin{table}[hbt!]
\caption{\label{tab:atomistic-deep-learning}Databases (DB) and software packages for applying DL methods for atomistic design. Here `k' and `mil' denote data points in thousands and millions respectively.}
\begin{minipage}{174pt}
%\caption{\label{tab:atomistic-deep-learning}Databases (DB) and software packages for applying DL methods for atomistic design. Here `k' and `mil' denote data points in thousands and millions respectively.}

\begin{tabular}{@{}llll@{}}
\toprule
\multicolumn{4}{c}{Databases} \\
\midrule
DB name & Datasize & Link & Ref\\
\midrule
JARVIS-DFT&  56k & \url{https://jarvis.nist.gov/jarvisdft/}   & \cite{choudhary2020joint}  \\
JARVIS-FF&  2.5k & \url{https://jarvis.nist.gov/jarvisff/}   & \cite{choudhary2020joint}  \\
MP&  144k & \url{https://materialsproject.org/}   & \cite{jain2013commentary}  \\
OQMD&  816k & \url{http://oqmd.org/}   & \cite{kirklin2015open}  \\
AFLOW&  3.5mil & \url{http://www.aflowlib.org/}   & \cite{curtarolo2012aflow}  \\
QM9&134k & \url{http://quantum-machine.org/datasets/} & \cite{ramakrishnan2014quantum}  \\
ANI&20mil& \url{https://github.com/isayev/ANI1_dataset} & \cite{smith2017ani}  \\
MD17&1mil& \url{http://quantum-machine.org/datasets} & \cite{chmiela2017machine}  \\
Tox21&760k & \url{https://tox21.gov/resources/}   & \cite{thomas2018us}  \\
CCCBDB&  2069 & \url{https://cccbdb.nist.gov/}   & \cite{russell2005nist}  \\
HOPV15&  350   & \url{https://doi.org/10.6084/m9.figshare.1610063} & \cite{lopez2016harvard}  \\
C2DB&  4000 & \url{https://cmr.fysik.dtu.dk/c2db/c2db.html}   & \cite{johnson2006nist}  \\
FreeSolv &  504 & \url{https://github.com/MobleyLab/FreeSolv}   & \cite{mobley2014freesolv}  \\
NOMAD&  11mil & \url{https://nomad-lab.eu/prod/rae/gui/search}   & \cite{draxl2018nomad}  \\
Open catalys\\
projectP&  1.2mil & \url{https://opencatalystproject.org}   & \cite{chanussot2021open}  \\
MCloud&  22mil & \url{https://www.materialscloud.org/home#statistics}   & \cite{talirz2020materials}  \\
CoreMOF&  163k   & \url{https://mof.tech.northwestern.edu/} & \cite{chung2019advances}  \\
QMOF& 22k &  \url{https://github.com/arosen93/QMOF}   & \cite{rosen2021machine}  \\
PDB & 183k &  \url{https://www.rcsb.org/}   & \cite{sussman1998protein}  \\
PDBBind  & 23k &  \url{http://www.pdbbind.org.cn/}   & \cite{wang2005pdbbind}  \\
MOAD   &  39k & \url{http://www.bindingmoad.org/}   & \cite{benson2007binding}  \\

%#########################################################%
\midrule
\multicolumn{3}{c}{Software packages} \\
\midrule
Model name & Applications& Link  & Ref\\
\midrule
ALIGNN &Mol, Sol & \url{https://github.com/usnistgov/alignn} & \cite{choudhary2021atomistic}\\
SchNetPack &Mol, Sol & \url{https://github.com/atomistic-machine-learning}    & \cite{schutt2018schnetpack}\\
CGCNN & Sol & \url{https://github.com/txie-93/cgcnn} & \cite{xie2018crystal}\\
MEGNet &Mol, Sol & \url{https://github.com/materialsvirtuallab/megnet} & \cite{chen2019graph}\\
DimeNet & Mol & \url{https://github.com/klicperajo/dimenet} & \cite{klicpera2020directional}\\
MPNN   & Mol & \url{https://github.com/priba/nmp_qc}  &  \cite{Gilmer2017}\\
ANI   & Mol & \url{https://github.com/isayev/ASE_ANI}     & \cite{smith2017ani}  \\
Amp  & Sol & \url{https://bitbucket.org/andrewpeterson/amp}    & \cite{khorshidi2016amp}  \\
TensorMol & Mol & \url{https://github.com/jparkhill/TensorMol}   & \cite{yao2018tensormol}  \\
PROPhet  & Sol & \url{https://github.com/biklooost/PROPhet}    & \cite{kolb2017discovering}  \\
DeepMD & Mol & \url{https://github.com/deepmodeling/deepmd-kit} & \cite{zhang2018deep,WANG2018178} \\
ænet& Sol & \url{https://github.com/atomisticnet/aenet} & \cite{ARTRITH2016135} \\
E3NN & Mol &\url{https://github.com/e3nn/e3nn}  & \cite{mario_geiger_2021_5006322} \\
Neural\\
fingerprint & Mol & \url{https://github.com/HIPS/neural-fingerprint}  & \cite{duvenaudConvolutionalNetworksGraphs2015} \\
DeepChemSt. & Mol & \url{https://github.com/MingCPU/DeepChemStable}  & \cite{LiDeepChemStable2019} \\
MoleculeNet & Mol, Sol &\url{https://github.com/deepchem/deepchem}  & \cite{wuMoleculeNetBenchmarkMolecular2018} \\
dgl-lifesci & Prot &\url{https://github.com/awslabs/dgl-lifesci}  & \cite{dgllife} \\
gnina & Prot &\url{https://github.com/gnina/gnina}  & \cite{mcnutt2021gnina} \\
\botrule
\end{tabular}
\end{minipage}
\end{table}

Table~\ref{tab:atomistic-deep-learning} we provide DL software packages used for atomistic materials design. The type of models includes general property (GP) predictors and interatomic force fields (FF). The models have been demonstrated in molecules (Mol), solid state materials (Sol) or proteins (Prot). For some force fields, high performance large scale implementations (LSI) that leverage paralleling computing exist. Some of these methods mainly used interatomic distances to build graphs while others use distances as well as bond angle information. Recently including bond angle within GNN has shown to drastically improve the performance with comparable computational timings.
 
 %See also: Software such as DeepMD, AeNet, and DeepMD are used for facilitating DL-based MD simulations.%

\subsubsection{Applications}
% interatomic potentials and dynamic simulation
% property prediction and screening
% thermodynamic stability prediction
% DFT hyperparameter selection / accelerated convergence / geometry optimization
% generative models and inverse design

% do we want to make the distinction here between local descriptor methods and GNNs?
% i.e. behler-parinello networks for graph neural networks?

 \paragraph{Force field development}

% predicting potential energy surfaces \cite{zhang2018deep}, adding corrections to computationally expensive n science 
%were made by \emph{ab-initio} calculations \cite{mcgibbon2017improving}, and alleviate computational expense of density functional theory (DFT), ab-initio molecular dynamics and other wave function methods  \cite{mills2017deep}.%

 The first application includes development of DL based force-fields (FF) \cite{smith2017ani,behler2011atom}/interatomic potentials. Some of the major advantages of such applications are that they are very fast (on the order of hundreds to thousands times \cite{wang2019deep}) for making predictions and solve the tenuous development of FFs, but the disadvantage is they still require a large dataset using computationally expensive methods to train. 
 
 Models such as Behler–Parrinello neural network (BPNN) and its variants \cite{behler2007generalized,Ko2021} are used for developing interatomic potentials that can be used for beyond just 0 K temperature and time dependent behavior using molecular dynamics simulations such as for nanoparticles \cite{weinreichproperties}. Such FF models have been developed for molecular systems such as water, methane and other organic molecules \cite{WANG2018178,Ko2021} as well as solids such as silicon \cite{behler2007generalized}, sodium \cite{eshet2010ab}, graphite \cite{khaliullin2010graphite} and titania ($TiO_2$) \cite{artrith2016implementation}. 
 
 While the above works are mainly based on NNs, there have also been development of graph neural network force field (GNNFF) framework \cite{park2021accurate,chmiela2018towards} that bypasses both computational bottlenecks. GNNFF can predict atomic forces directly using automatically extracted structural features that are not only translationally-invariant, but rotationally-covariant to the coordinate space of the atomic positions. In addition to development of pure NN based FFs, there have also been recent developments of combining traditional FFs such as bond-order potentials with NNs and ReaxFF with message passing neural network (MPNN) that can help mitigate the NNs issue for extrapolation \cite{pun2019physically,xue2021reaxff}.
 
 %https://pubs.acs.org/doi/abs/10.1021/acs.jpcc.0c00559%
 \paragraph{Direct property prediction from atomistic configurations}
 DL methods can be used to to establish structure-property relationship between atomic structure and their properties with high accuracy \cite{kearnes2016molecular,Gilmer2017}.
 Models such as SchNet, crystal graph convolutional neural network (CGCNN), improved crystal graph convolutional neural network (iCGCNN), directional message passing neural network (DimeNet), atomistic line graph neural network (ALIGNN) and materials graph neural network (MEGNet) shown in Table~\ref{tab:atomistic-deep-learning} have been used to predict up to 50 properties of crystalline and molecular materials. These property datasets are usually obtained from ab-initio calculations. A schematic of such models shown in Fig. 2. While SchNet, CGCNN, MEGNet are primarily based on atomic distances only, iCGCNN, DimeNet, and ALIGNN models capture many body interactions using GCNN.
 
 Some of these properties include formation energies, electronic bandgaps, solar-cell efficiency, topological spin-orbit spillage, dielectric constants, piezoelectric constants, 2D exfoliation energies, electric field gradients, elastic modulus, Seebeck coefficients, power factors,  carrier effective masses, highest occupied molecular orbital, lowest unoccupied molecular orbital, energy gap, zero-point vibrational energy, dipole moment, isotropic polarizability, electronic spatial extent, internal energy. 
 
 For instance, the current state of the art mean absolute error for formation energy for solids and internal energy for molecules at 0 K are 0.022 eV/atom and 0.002 eV as obtained by the ALIGNN model~\cite{choudhary2021atomistic}. DL is also heavily being used for predicting catalytic behavior of materials such as the Open Catalyst Project \cite{zitnick2020introduction} which is driven by the DL methods materials design. There is an ongoing effort to continuously improve the models. Usually energy based models such as formation and total energies are more accurate than electronic property based models such as bandgaps and power factors. 
 
 In addition to molecules and solids, property predictions models have also been used for bio-materials such as proteins, which can be viewed as a large molecule. There have been several efforts for predicting protein based properties such as binding affinity \cite{dgllife} and docking predictions \cite{mcnutt2021gnina}.

There have been also several applications for identifying reasonable chemical space using DL methods such as autoencoders \cite{jin2018junction}, reinforcement learning \cite{olivecrona2017molecular,you2018graph,putin2018reinforced} for inverse materials design. Inverse materials design with techniques such as GAN deals with finding chemical compounds with suitable properties and act as complementary to forward prediction models. While such concepts have been widely applied to molecular systems, \cite{sanchez2017optimizing}, recently these methods have been applied to solids as well \cite{nouira2018crystalgan,Long2021,Noh2019,Kim2020,long2021inverse}. 

%[I would suggest Chi add a paragraph on multi-fidelity graph networks that deals with the small data problem.]

\paragraph{Fast materials screening}
DFT based high-throughput methods are usually limited to few thousands of compounds and takes a long time for calculations, DL based methods can aid this process and allow much faster predictions. DL based property prediction models mentioned above can be used for pre-screening chemical compounds. Hence, DL based tools can be viewed as a pre-screening tool for traditional methods such as DFT. For example, Xie et al. used CGCNN model to screen stable perovskite materials \cite{xie2018crystal} as well hierarchical visualization of materials
space \cite{xie2018hierarchical}.
Park et al. \cite{park2020developing} used iCGCNN to screen $ThCr_2Si_2$-type materials. Lugier et al used DL methods to predict thermoelectric properties \cite {laugier2018predicting}. Rosen et al.
\cite{rosen2021machine} used graph neural network models to predict the bandgaps of metal organic frameworks.
DL for molecular materials have been used to predict technologically important properties such as aqueous solubility \cite{lusci2013deep} and toxicity \cite{xu2015deep}.

\begin{figure}
    \centering
    \includegraphics[trim={0 .5cm 0 1cm},clip,width=1.0\textwidth]{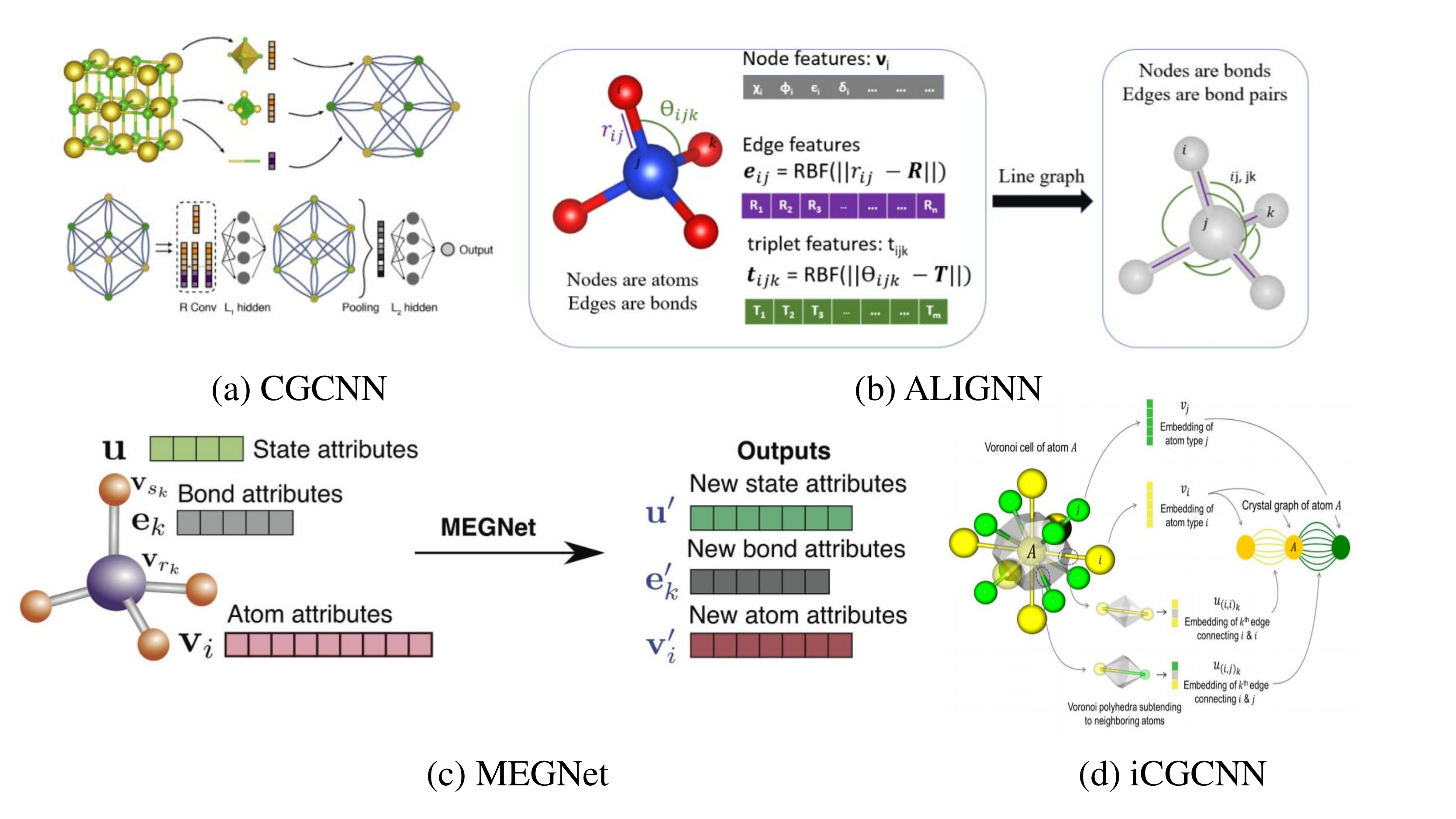}
    \caption{Schematic representations of an atomic structure as a graph. a) CGCNN model in which  crystals are
converted to graphs with nodes representing atoms in the unit cell and edges representing atom connections. Nodes and edges are characterized by vectors corresponding to the atoms and bonds in the crystal, respectively [Reprinted with permission from ref. \cite{xie2018crystal} Copyright 2019 American Physical Society], b) ALIGNN \cite{choudhary2021atomistic} model in which the convolution layer alternates between message passing on the bond graph and its bond-angle line graph c) MEGNet in which the initial graph is represented by the set of atomic attributes, bond attributes and global state attributes  [Reprinted with permission from ref. \cite{chen2019graph} Copyright 2019 American Chemical Society] model, d) iCGCNN model in which multiple edges connect a node to neighboring nodes to show the number of Voronoi neighbors [Reprinted with permission from ref. \cite{park2020developing} Copyright 2019 American Physical Society]}.
\end{figure}

 % there's an aspect to this work about predicting adsorption sites, ligand geometry, etc right?
 % TODO: review Noa Marom's recent work on molecular crystals and see if anything fits?

% \paragraph{Complimentary methods to atomistic approach}
It should be noted that the full atomistic representations and the associated DL models are only possible if the crystal structure and atom positions are available. In practice, the precise atom positions are only available from DFT structural relaxations or experiments, and are one of the goals for materials discovery instead of the starting point. Hence, alternative methods have been proposed to bypass the necessity for atom positions in building DL models. For example, Jain and Bligaard \cite{jainAtomicpositionIndependentDescriptor2018} proposed the atomic position independent descriptors and used a CNN model to learn energies of crystals. Such descriptors include information only on the symmetry information (e.g., spacegroup and Wyckoff position). In principle, the method can be applied universally in all crystals. Nevertheless, the model errors tend to be much higher than graph-based models. Similar coarse-grained representation using Wyckoff representation was also used by Goodall et al.\cite{goodall2021rapid}. Alternatively, Zuo et al.\cite{zuoAcceleratingMaterialsDiscovery2021} started from the hypothetical structures without precise atom positions, and used a Bayesian optimization method coupled with a MEGNet energy model as energy evaluator to perform direct structural relaxation. The application of the developed Bayesian optimization with symmetry relaxation (BOWSR) algorithm successfully discovered ReWB (Pca2$_1$) and MoWC$_2$ (P6$_3$/mmc) hard materials, which were then experimentally synthesized.

% some categories of tasks to sort through?
% Screening, interpretability, binding affinity, drug discovery, interatomic potential, catalyst, stability prediction, beyond just emulating DFT, other high level DFT, generative models for molecules and crystals (e.g. gan and vae), drug binding affinity and docking prediction, inverse design, [generate molecules and solids], reaction prediction and link prediction, reaction mechanism, NEB, Autonomous

%- https://www.biorxiv.org/content/10.1101/2021.01.07.425790v1.full
%- Isayev graphlet representation

\subsection{Chemical formula and segment representations}\label{sec:stoichiometric}
% property prediction and screening
% thermodynamic stability prediction
% structure prediction
% generative models and inverse design
One of the earliest applications for DL included SMILES for molecules, elemental fractions and chemical descriptors for solids and sequence of protein names as descriptors. Such descriptors lack explicit inclusion of atomic structure information but are still useful for various pre-screening applications for both theoretical and experimental data. 

%Murat paper on 89-dimensional phase diagram \cite{aykolstabilitynetwork}%.
% not deep learning, but somebody should do some

\subsubsection{SMILES and fragment representation}

The simplified molecular-input line-entry system (SMILES) is a method to represent elemental and bonding for molecular structures using short American Standard Code for Information Interchange (ASCII) strings. SMILES can express structural differences including the chirality of compounds making it more useful than simply chemical formula. A SMILES string is a simple grid-like (1-D grid) structure that can represent molecular sequences such as DNA, macromolecules/polymers, protein sequences also \cite{lin2019bigsmiles,tyagi2015cancerppd}. In addition to the chemical constituents as in chemical formula, bondings (such as double and triple bondings) are represented by special symbols (such as '=' and '\#'). The presence of a branch point indicated using a left-hand bracket “(” while the right-hand bracket “)” indicates that all the atoms in that branch have been taken into account. SMILES strings are represented as a distributed representation termed a SMILES feature matrix (as a sparse matrix), and then we can apply DL to the matrix similar to image data. The length of the SMILES matrix is generally kept fixed (such as 400) during training and in addition to the SMILES multiple elemental attributes and bonding attributes (such as chirality, aromaticity) can be used. Key DL tasks for molecules include a) novel molecule design, b) molecule screening.

Novel molecules with target properties can designed using VAE, GAN and RNN based methods \cite{krenn2020self,lim2018molecular,krasnov2021transformer}. These DL generated molecules might not be physically valid, but the goal is to train the model to learn the patterns in SMILES strings such that the output resembles valid molecules. Then chemical intuitions can be further used to screen the molecules. DL for SMILES can also be used for molecularscreening such as to predict molecular toxicity. Some of the common SMILES datasets are: ZINC \cite{irwin2012zinc}, Tox21 \cite{dix2007toxcast} and PubChem \cite{kim2019pubchem}.

Due to the limitations to enforce the generation of valid molecular structures from SMILES, fragment based models are developed such as DeepFrag and DeepFrag-K \cite{hirohara2018convolutional,gomez2018automatic}. In fragment based models, a ligand/receptor complex is removed and then a DL model is trained to predict the most suitable fragment substituent. A set of useful tools for SMILES and fragment representations are provided in Table 2.

\subsubsection{Chemical formula representation}

There are several ways of using the chemical formula based representations for building ML/DL models, beginning with a simple vector of raw elemental fractions \cite{liu2016deep,jha2018elemnet} or of weight percentages of alloying compositions \cite{agrawal2014exploration,agrawal2016fatigue,agrawal2018online, agrawal2019martensite}, as well as more sophisticated hand-crafted descriptors or physical attributes to add known chemistry knowledge (e.g. electronegativity, valency, etc. of constituent elements) to the feature representations \cite{meredig2014combinatorial,agrawal2016formation,furmanchuk2016predictive,furmanchuk2018prediction,ward2016general,ward2018matminer}. Statistical and mathematical operations such as average, max, min , median, mode, and exponentiation can be carried out on elemental properties of the constituent elements to get a set of descriptors for a given compound. The number of such composition-based features can range from a few dozens to few hundreds. One of the commonly used representations that has been shown to work for a variety of different use-cases is the materials agnostic platform for informatics and exploration (MagPie) \cite{ward2016general}. All these composition-based representations can be used with both traditional ML methods such as Random Forest as well as DL. 

It is relevant to note that ElemNet \cite{jha2018elemnet}, which is a 17-layer neural network composed of fully-connected layers and uses only raw elemental fractions as input, was found to significantly outperform traditional ML methods such as Random Forest, even when they were allowed to use more sophisticated physical attributes based on MagPie as input. Although no periodic table information was provided to the model, it was found to self-learn some interesting chemistry, like groups (element similarity) and charge balance (element interaction), and was also able to predict phase diagrams on unseen materials systems, underscoring the power of DL for representation learning directly from raw inputs without explicit feature extraction. Further increasing the depth of the network was found to adversely affect the model accuracy due to the vanishing gradient problem. To address this issue, Jha et al. \cite{jha2019irnet} developed IRNet, which uses individual residual learning to allow a smoother flow of gradients and enable deeper learning for cases where big data is available. IRNet models were tested on a variety of big and small materials datasets, such as OQMD, AFLOW, Materials Project, JARVIS, using different vector-based materials representations (element fractions, MagPie, structural) and were found to not only successfully alleviate the vanishing gradient problem and enable deeper learning, but also lead to significantly better model accuracy as compared to plain deep neural networks and traditional ML techniques for a given input materials representation in the presence of big data \cite{jha2021enabling}. Further, graph based methods such as Roost~\cite{goodall2020predicting} have also been developed which can outperform many similar techniques.

Such methods have been used for diverse DFT datasets mentioned above in Table 1 as well as experimental datasets such as SuperCon ~\cite{supercon,stanev2018machine} for quick pre-screening applications. In terms of applications, they have have been applied for predicting properties such as formation energy \cite{jha2018elemnet}, band gap and magnetization \cite{jha2019irnet}, superconducting temperatures \cite{stanev2018machine}, bulk and shear modulus \cite{jha2021enabling}. They have also been used for transfer learning across datasets for enhanced predictive accuracy on small data \cite{jha2019enhancing}, even for different source and target properties \cite{gupta2021cross}. 

There have been libraries of such descriptors developed such as MatMiner ~\cite{ward2018matminer} and DScribe ~\cite{himanen2020dscribe}. Some examples of such models are given in Table~\ref{tab:stoichiometric}. Such representations are especially useful for experimental dataset such as superconducting material dataset where actual atomic structure is not known. However, these representations cannot distinguish different polymorphs of a system with different point groups and space groups. It has been recently shown that although composition-based representations can help build ML/DL models to predict some properties like formation energy with a remarkable accuracy, it does not necessarily translate to accurate predictions of other properties such as stability, when compared to DFT's own accuracy \cite{bartel2020critical}.

\begin{table}[hbt!]
\caption{Software packages for applying DL methods for chemical formula, SMILES and fragment representations.}\label{tab:stoichiometric}%
\begin{minipage}{174pt}
%\caption{Software packages for applying DL methods for chemical formula, SMILES and fragment representations}\label{tab:stoichiometric}%
\begin{tabular}{@{}llll@{}}
\toprule
\multicolumn{3}{c}{Chemical formula} \\
\midrule
Model name & Link  & Reference\\
\midrule
MatMiner & \url{https://github.com/hackingmaterials/matminer} & \cite{ward2018matminer} \\
MagPie   & \url{https://bitbucket.org/wolverton/magpie}  & \cite{ward2016general}  \\
DScribe   & \url{https://github.com/SINGROUP/dscribe}  & \cite{himanen2020dscribe}  \\
ElemNet   & \url{https://github.com/NU-CUCIS/ElemNet}    & \cite{jha2018elemnet}  \\
IRNet   & \url{https://github.com/NU-CUCIS/IRNet}   & \cite{jha2019irnet,jha2021enabling}   \\
Roost  & \url{https://github.com/CompRhys/roost}  & \cite{goodall2020predicting}   \\
CrabNet  & \url{https://github.com/anthony-wang/CrabNet}   & \cite{Wang2021crabnet}  \\
CFID-Chem & \url{https://github.com/usnistgov/jarvis/} & \cite{choudhary2018machine} \\
Atom2vec & \url{https://github.com/idocx/Atom2Vec} & \cite{zhou2018learning} \\

\midrule
\multicolumn{3}{c}{SMILES and fragments} \\
\midrule
DeepSMILES  & \url{https://github.com/baoilleach/deepsmiles}  & \cite{oboyle_dalke_2018} \\
ChemicalVAE  & \url{https://github.com/aspuru-guzik-group/chemical_vae} & \cite{chemicalvae}  \\
CVAE & \url{https://github.com/jaechanglim/CVAE} & \cite{lim2018molecular}\\
DeepChem & \url{https://github.com/deepchem/deepchem} & \cite{wuMoleculeNetBenchmarkMolecular2018} \\
DeepFRAG  & \url{https://git.durrantlab.pitt.edu/jdurrant/deepfrag/}  &  \cite{green2021deepfrag}\\
DeepFRAG-k  & \url{https://github.com/yaohangli/DeepFragK/}  &  \cite{elhefnawy_li_wang_li_2020}\\
CheMixNet & \url{https://github.com/NU-CUCIS/CheMixNet} & \cite{paul2018chemixnet}\\
SINet & \url{https://github.com/NU-CUCIS/SINet} & \cite{paul2019transfer}\\

\botrule
\end{tabular}
\end{minipage}
\end{table}

%https://chemrxiv.org/engage/api-gateway/chemrxiv/assets/orp/resource/item/60c74849567dfefc91ec49a1/original/is-domain-knowledge-necessary-for-machine-learning-materials-properties.pdf

\subsection{Spectral models}\label{sec:spectral}

When electromagnetic radiation hits materials, the interaction between the radiation and matter measured as a function of the wavelength or frequency of the radiation produces a spectroscopic signal. By studying spectroscopy, researchers can gain insights into the materials’ composition, structural, and dynamic properties. Spectroscopic techniques are foundational in materials characterization. For instance, X-ray diffraction (XRD) has been used to characterize the crystal structure of materials for more than a century.
Spectroscopic analysis can involve fitting quantitative physical models (for example, Rietveld refinement) or more empirical approaches such as fitting linear combinations of reference spectra, such as with x-ray absorption near edge spectroscopy (XANES). Both approaches require a high degree of researcher expertise through careful design of experiments; specification, revision, and iterative fitting of physical models; or the availability of template spectra of known materials. In recent years, with the advances in high-throughput experiments and computational data, spectroscopic data has multiplied, giving opportunities for researchers to learn from the data and potentially displace the conventional methods in analyzing such data. This section covers emerging DL applications in various modes of spectroscopic data analysis, aiming to offer practice examples and insights. Some of the applications are shown in Fig.3.

\subsubsection{Databases and software libraries}

\begin{table}[hbt!]
\caption{Databases and software packages for applying DL methods for spectra data.}\label{tab:atomistic-datasets}%
\begin{minipage}{174pt}

 \begin{tabular}{@{}llll@{}}

\toprule
 & & Databases &\\
\midrule
DB name & Datasize & Link & Ref\\
\midrule
MP XAS-DB  &  490k   & \url{https://materialsproject.org/} & \cite{mathew2018high,chen2021database}  \\
JV Dielectric\\ function  &  16k   & \url{http://jarvis.nist.gov/jarvisdft} & \cite{choudhary2018computational}  \\
JV Infrared  &  5k   & \url{http://jarvis.nist.gov/jarvisdft} & \cite{choudhary2020high}  \\
RRUFF  &  3527   & \url{https://rruff.info} & \cite{lafuente20151}  \\
ICDD XRD  &  108k   & \url{https://www.icdd.com/pdf-product-summary/ } & \cite{wong2001jcpds}  \\
ICSD XRD  &  150k   & \url{https://icsd.nist.gov/} & \cite{belsky2002new}  \\
COD XRD  & 480k & \url{http://www.crystallography.net/cod/} & \cite{Grazulis2009} \\
MP XRD  &  140k   & \url{https://materialsproject.org/} & \cite{jain2013commentary}  \\
JV XRD  &  60k   & \url{https://jarvis.nist.gov/jarvisdft/} & \cite{choudhary2020joint}  \\
MPContribs & - & \url{https://mpcontribs.org/} & 
\cite{mpcontribs2019}\\
Raman \\OpenDB  &  1k   & \url{https://solsa.crystallography.net/rod/index.php } & \cite{el2019raman}  \\
Chem. Web  &  1k   & \url{https://webbook.nist.gov/chemistry/ } & \cite{linstrom2001nist} \\
PDFitc XPD & - & \url{https://pdfitc.org} & \cite{yang;aca21} \\
SDBS  &  35k  & \url{http://sdbs.riodb.aist.go.jp/sdbs/cgi-bin/cre_index.cgi} & \cite{saito2006spectral}  \\
NMRShiftDB  &  44k  & \url{https://nmrshiftdb.nmr.uni-koeln.de/} & \cite{steinbeck2003nmrshiftdb}  \\
SpectraBase  &  -  & \url{https://spectrabase.com/} & \cite{steinbeck2003nmrshiftdb}  \\
SOP  &  325  & \url{https://soprano.kikirpa.be/index.php?lib=sop} & \cite{fremout2012identification}  \\
HTEM  &  140k  & \url{https://htem.nrel.gov/} & \cite{zakutayev2018open}  \\
%###############################################%
\midrule
 & & Software packages&\\
\midrule
Software name & Type & Link   & Ref\\
\midrule
DOSNet   & Sol &  \url{https://github.com/vxfung/DOSnet}   & \cite{fung2021machine}  \\
PCA-CGCNN   & Sol & \url{https://github.com/kihoon-bang/PCA-CGCNN}   & \cite{bang2021accelerated}  \\
autoXRD   & Sol &  \url{https://github.com/PV-Lab/autoXRD}   & \cite{oviedoFastInterpretableClassification2019}  \\
PDFitc XPD & Sol & \url{https://pdfitc.org} & \cite{yang;aca21} \\

\botrule
\end{tabular}
\end{minipage}
\end{table}

Currently, large-scale and element-diverse spectral data mainly exist in computational databases. For example, in Ref. \cite{choudhary2020high}, the authors calculated the infrared spectra, piezoelectric tensor, Born effective charge tensor, and dielectric response as a part of the JARVIS-DFT DFPT database. The Materials Project has established the largest computational X-ray absorption database (XASDb), covering the K-edge X-ray near-edge fine structure (XANES) \cite{zhengAutomatedGenerationEnsemblelearned2018,mathew2018high} and the L-edge XANES \cite{chen2021database} of a large number of material structures. The database currently hosts more than 400000 K-edge XANES site-wise spectra and ~90000 L-edge XANES site-wise spectra of many compounds in the Materials Project. There is considerably fewer experimental XAS spectra, being on the order of hundreds, as seen in the EELSDb and the XASLib. Collecting large experimental spectra databases that cover a wide range of elements is a challenging task. Collective efforts have been focusing on curating data extracted from different sources, as found in the RRUFF Raman, XRD and chemistry database \cite{lafuente20151}, the open Raman database \cite{el2019raman}, and the SOP spectra library \cite{fremout2012identification}. However, data consistency is not guaranteed.  It is also now possible for contributors to share experimental data in a Materials Project curated database, MPContribs \cite{mpcontribs2019}. This database is supported by the US Department of Energy (DOE) providing some expectation of persistence. Entries can be kept private or published and are linked to the main materials project computational databases.  There is an ongoing effort to capture data from DOE funded synchrotron light sources (https://lightsources.materialsproject.org/) into MPContribs in the future.

Recent advances in sources, detectors and experimental instrumentation have made high-throughput measurements of experimental spectra possible, giving rise to new possibilities for spectral data generation and modeling. Such examples include the HTEM database \cite{zakutayev2018open} that contains 50000 optical absorption spectra, the UV-Vis database of ~180000 samples from the Joint Center for Artificial Photosynthesis.  Some of the common spectra databases for spectra data are shown in Table \ref{tab:atomistic-datasets}.  There are beginning to appear cloud-based software as a service platforms for high throughput data analysis, for example, pair-distribution function (PDF) in the cloud (https://pdfitc.org) \cite{yang;aca21} which are backed by structured databases, where data can be kept private or made public.  This transition to the cloud from data analysis software installed and run locally on a user's computer will facilitate the sharing and reuse of data by the community.

\subsubsection{Applications}
Due to the widespread deployment of XRD across many materials technologies, XRD spectra became one of the first test grounds for DL models. Phase identification from XRD can be mapped into a classification task (assuming all phases are known) or an unsupervised clustering task. Multi-phase diffraction data Unlike the traditional analysis of XRD data, where the spectra are treated as convolved, discrete peak positions and intensities, DL methods treat the data as an continuous pattern similar to an image. Unfortunately, a significant number of experimental XRD data-sets in one place are not readily available at the moment. Nevertheless, extensive, high-quality crystal structure data makes creating simulated XRD trivial.

\begin{figure}
    \centering
    \includegraphics[width=1.0\textwidth]{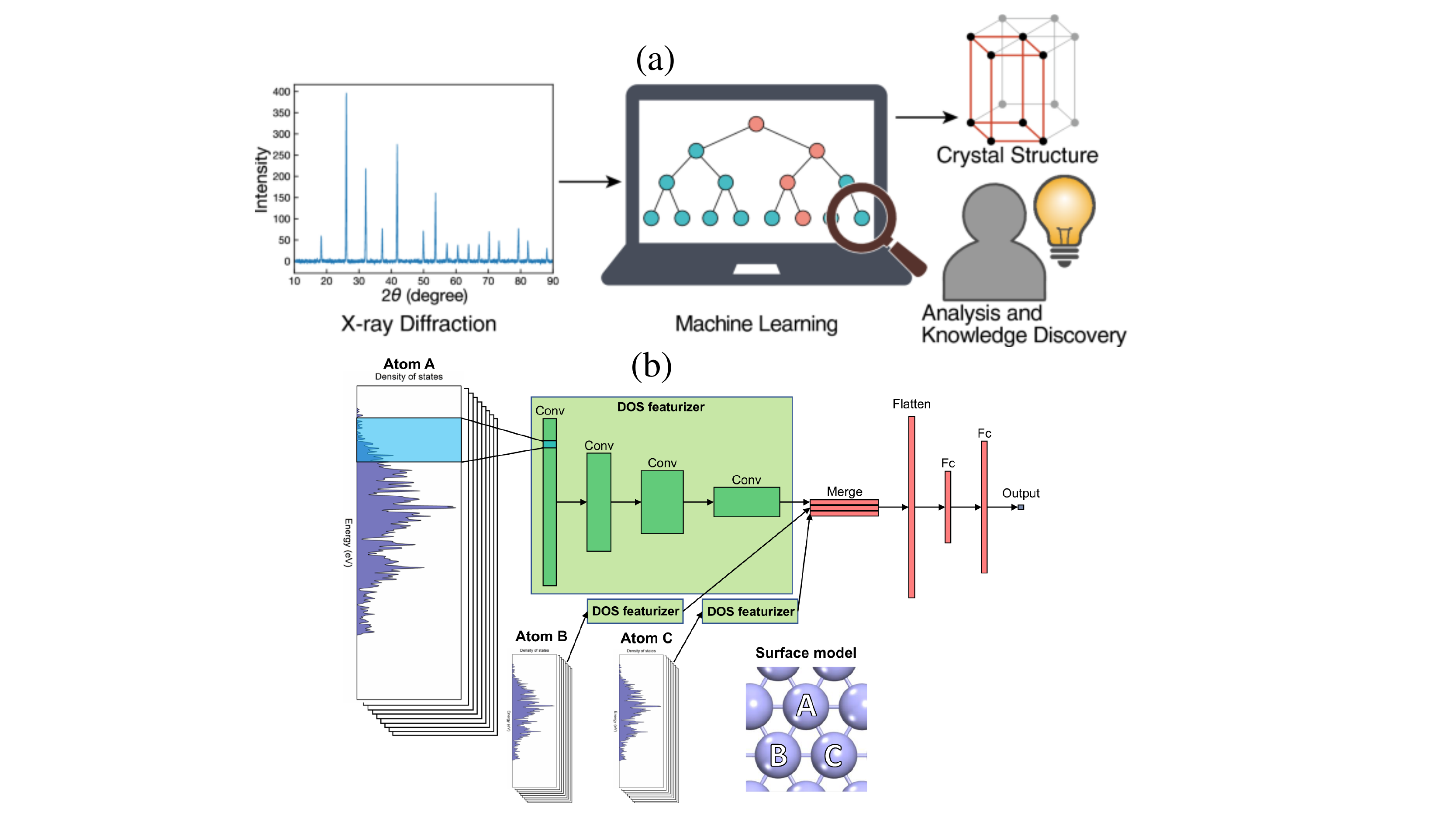}
    \caption{Example applications of deep-learning for spectral data. a) Predicting structure information from the X-ray diffraction \cite{suzukiSymmetryPredictionKnowledge2020},  Reprinted according to the terms of the CC-BY license.\cite{suzukiSymmetryPredictionKnowledge2020} Copyright 2020. b) Predicting catalysis properties from computational electronic density of states data. Reprinted according to the terms of the CC-BY license.\cite{fungMachineLearnedFeatures2021}. Copyright 2021.}
    \label{fig:spectra_model}
\end{figure}

Park et al. \cite{parkClassificationCrystalStructure2017} calculated 150000 XRD patterns from the Inorganic Crystal Structure Database (ICSD) structural database \cite{hellenbrandtInorganicCrystalStructure2004} and then used CNN models to predict structural information from the simulated XRD patterns. The accuracies of the CNN models reached 81.14 \%, 83.83 \%, and 94.99 \% for space-group, extinction-group, and crystal-system classifications, respectively.

Liu et al. \cite{liu2019using} obtained similar accuracies by using a CNN for classifying atomic pair distribution function (PDF) data into space groups. The PDF is obtained by Fourier transforming XRD into real-space and is particularly useful for studying the local and nano-scale structure of materials.  In the case of the PDF, models were trained, validated and tested on simulated data from the ICSD.  However, the trained model showed excellent performance when it was given experimental data, something that can be a challenge in XRD data because of the different resolutions and line-shapes of the diffraction data depending on specifics of the sample and experimental conditions.  The PDF seems to be more robust against these aspects.

Similarly, Zaloga et al. \cite{zalogaCrystalSymmetryClassification2020} also used the ICSD database for XRD pattern generation and CNN models to classify crystals. The models achieved 90.02 \% and 79.82 \% accuracy for crystal systems and space groups, respectively. 

It should be noted that the ICSD database contains many duplicates, and such duplicates should be filtered out to avoid information leakage. There is also a large difference in the number of structures represented in each space group (the label) in the database resulting in data normalization challenges.

Lee et al. \cite{leeDeeplearningTechniquePhase2020} developed a CNN model for phase identification from samples consisting of a mixture of several phases in a limited chemical space relevant for battery materials. The training data are mixed patterns consisting of 1785405 synthetic XRD patterns from the Sr-Li-Al-O phase space. The resulting CNN can not only identify the phases but also predict the compound fraction in the mixture. A similar CNN  was utilized by Wang et al. \cite{wangRapidIdentificationXray2020} for fast identification of metal-organic frameworks (MOFs), where experimental spectral noise was extracted and then synthesized into the theoretical XRD for training data augmentation.

An alternative idea was proposed by Dong et al. \cite{dongDeepConvolutionalNeural2021}, where instead of recognizing only phases from the CNN, a proposed ``parameter quantification network" (PQ-Net) was able to extract physico-chemical information. The PQ-Net yields accurate predictions for scale factors, crystallite size, and lattice parameters for simulated and experimental XRD spectra. The work by Aguiar et al. \cite{aguiarCrystallographicPredictionDiffraction2020} took a step further and proposed a modular neural network architecture that enables the combination of diffraction patterns and chemistry data and provided a ranked list of predictions. The ranked list predictions provides user flexibility and overcomes some aspects of overconfidence in model predictions. 
In practical applications, AI-driven XRD identification can be beneficial for high-throughput materials discovery, as shown by Maffettone et al. \cite{maffettoneCrystallographyCompanionAgent2021a} In their work, an ensemble of fifty CNN models was trained on synthetic data reproducing experimental variations (missing peaks, broadening, peaking shifting, noises). The model ensemble is capable of predicting the probability of each category label. A similar data augmentation idea was adopted by Oviedo et al. \cite{oviedoFastInterpretableClassification2019}, where experimental XRD data for 115 thin-film metal-halides were measured, and CNN models trained on the augmented XRD data achieved accuracies of 93 \% and 89 \% for classifying dimensionality and space group, respectively. 

Although not a DL method, an unsupervised machine learning approach, non-negative matrix factorization (NMF), is showing great promise for yielding chemically relevant XRD spectra from time- or spatially-dependent sets of diffraction patterns.  NMF is closely related to principle component analysis in that it takes a set of patterns as a matrix and then compresses the data by reducing the dimensionality by finding the most important components.  In NMF a constraint is applied that all the components and their weights must be strictly positive.  This often corresponds to a real physical situation (for example, spectra tend to be positive, as are the weights of chemical constituents).  As a result we are finding that the mathematical decomposition often results in interpretable, physically meaningful, components and weights, as shown by Liu et al. for PDF data \cite{liu;jac21}.  An extension of this showed that in a spatially resolved study, NMF could be used to extract chemically resolved differential PDFs (similar to the information in EXAFS) from non-chemically resolve PDF measurements \cite{rakit;arxiv21}.  NMF is very quick and easy to apply and can be applied to just about any set of spectra.  It is likely to become widely used and is being implemented in the PDFitc.org website to make it more accessible to potential users.

Other than XRD, the XAS, Raman, and infrared spectra, also contain rich structure-dependent spectroscopic information about the material. Unlike XRD, where relatively simple theories and equations exist to relate structures to the spectral patterns, the relationships between general spectra and structures are somewhat illusive. This difficulty has created a higher demand for machine learning models to learn structural information from other spectra.  

For instance, the case of X-ray absorption spectroscopy (XAS), including the X-ray absorption near-edge spectroscopy (XANES) and extended X-ray absorption fine structure (EXAFS), is usually used to analyze the structural information on an atomic level. However, the high signal-to-noise XANES region has no equation for data fitting. DL modeling of XAS data is fascinating and offers unprecedented insights. Timoshenko et al. used neural networks to predict the coordination numbers of Pt \cite{timoshenkoSupervisedMachineLearningBasedDetermination2017} and Cu \cite{timoshenkosubnano2018} in nanoclusters from the XANES. Aside from the high accuracies, the neural network also offers high prediction speed and new opportunities for quantitative XANES analysis. Timoshenko et al. \cite{timoshenkoNeuralNetworkApproach2018} further carried out a novel analysis of EXAFS using DL. Although EXAFS analysis has an explicit equation to fit, the study is limited to the first few coordination shells and on relatively ordered materials.  Timoshenko et al.  \cite{timoshenkoNeuralNetworkApproach2018} first transformed the EXAFS data into 2D maps with a wavelet transform and then supplied the 2D data to a neural network model. The model can instantly predict relatively long-range radial distribution functions, offering \emph{in situ} local structure analysis of materials. The advent of high-throughput XAS databases has recently unveiled more possibilities for machine learning models to be deployed using XAS data. For example, Zheng et al. \cite{zhengAutomatedGenerationEnsemblelearned2018} used an ensemble learning method to match and fast search new spectra in the XASDb. Later, the same authors showed that random forest models outperform DL models such as MLPs or CNNs in predicting atomic environment labels from the XANES spectra directly \cite{zhengRandomForestModels2020}. Similar approaches were also adopted by Torrisi et al. \cite{torrisiRandomForestMachine2020} In practical applications, Andrejevic et al. \cite{andrejevic2020machine} used the XASDb data together with the topological materials database and constructed CNN models to classify the topology of materials from the XANES and symmetry group inputs. The model correctly predicted 81 \% topological and 80 \% trivial cases and achieved 90 \% accuracy in material classes that contain certain elements. 

Raman, infrared, and other vibrational spectroscopies provide structural fingerprints and are usually used to discriminate and estimate the concentration of components in a mixture. For example, Madden et al. \cite{maddenMachineLearningMethods2003} have used neural network models to predict the concentration of illicit materials in a mixture using the Raman spectra. Interestingly, several groups have independently found that DL models outperform chemometrics analysis in vibrational spectroscopies \cite{conroyQualitativeQuantitativeAnalysis2005, acquarelliConvolutionalNeuralNetworks2017}. For learning vibrational spectra, the number of training spectra is usually less than or on the order of the number of features (intensity points), and the models can easily overfit. Hence, dimensional reduction strategies are commonly used to compress the information dimension using, for example, principal component analysis (PCA) \cite{oconnellClassificationTargetAnalyte2005, zhaoQualitativeIdentificationTea2006}. DL approaches do not have such concerns and offer elegant and unified solutions. For example, Liu et al.\cite{liuDeepConvolutionalNeural2017} applied CNN models to the Raman spectra in the RRUFF spectral database and show that CNN models outperform classical machine learning models such as SVM in classification tasks. More DL applications in vibrational spectral analysis can be found in a recent review by Yang et al. \cite{yangDeepLearningVibrational2019}

Although most current DL work focuses on the inverse problem, i.e., predicting structural information from the spectra, some innovative approaches also solve the forward problems by predicting the spectra from the structure. In this case, the spectroscopy data can be viewed simply as a high-dimensional material property of the structure. This is most common in molecular science, where predicting the infrared spectra \cite{selzer_gasteiger_thomas_salzer_2000}, molecular excitation spectra \cite{ghosh_rinke_2019}, is of particular interest. In the early 2000s, Selzer et al. \cite{selzer_gasteiger_thomas_salzer_2000} and Kostka et al. \cite{kostka_selzer_gasteiger_2001} attempted predicting the infrared spectra directly from the molecular structural descriptors using neural networks. Non-DL models can also be used to perform such tasks to a reasonable accuracy \cite{mahmoudLearningElectronicDensity2020}. For DL models, Chen et al. \cite{chenDirectPredictionPhonon2021} used a Euclidean neural network (E(3)NN) to predict the phonon density of state (DOS) spectra from atom positions and element types. The E(3)NN model captures symmetries of the crystal structures, with no need to perform data augmentation to achieve target invariances. Hence the E(3)NN model is extremely data-efficient and can give reliable DOS spectra prediction and heat capacity using relatively sparse data of 1200 calculation results on 65 elements. A similar idea was also used to predict the XAS spectra. Carbone et al. \cite{carboneMachineLearningXRayAbsorption2020a} used a message passing neural network (MPNN) to predict the O and N K-edge XANES spectra from the molecular structures in the QM9 database \cite{ramakrishnan2014quantum}. The training XANES data were generated using the FEFF package \cite{rehrParameterfreeCalculationsXray2010}. The trained MPNN model reproduced all prominent peaks in the predicted XANES, and 90 \% of the predicted peaks are within 1 eV of the FEFF calculations. Similarly, Rankine et al. \cite{rankineDeepNeuralNetwork2020} started from the two-body radial distribution function (RDC) and used a deep neural network model to predict the Fe K-edge XANES spectra for arbitrary local environments.

In addition to learn the structure-spectra or spectra-structure relationships, a few works have also explored the possibility of relating spectra to other material properties in a non-trivial way. The DOSnet proposed by Fung et al. \cite{fungMachineLearnedFeatures2021} (Figure \ref{fig:spectra_model}b) uses the electronic DOS spectra calculated from DFT as inputs to a CNN model to predict the adsorption energies of H, C, N, O, S and their hydrogenated counterparts, CH, CH$_2$, CH$_3$, NH, OH, and SH,  on bimetallic alloy surfaces. This approach extends the previous d-band theory  \cite{hammer_norskov_2000}, where only the d-band center, a scalar, was used to correlate with the adsorption energy on transition metals. Stein et al.   \cite{stein_soedarmadji_newhouse_guevarra_gregoire_2019} tried to learn the mapping between the image and the UV-vis spectrum of the material using the conditional variational encoder (cVAE) with neural network models as the backbone. Such models can generate the UV-vis spectrum directly from a simple material image, offering much faster material characterizations.

\subsection{Image based models}\label{sec:image}
Computer vision is often credited as the precipitating the current wave of mainstream DL applications a decade ago \cite{krizhevsky2012imagenet}.
Naturally, materials researchers have developed a broad portfolio of applications of computer vision for accelerating and improving image-based material characterization techniques. 
High-level microscopy vision tasks can be organized as follows:
\begin{itemize}
    \item image classification (and material property regression)
    \item auto-tuning experimental imaging hyperparameters
    \item pixel-wise learning (e.g. semantic segmentation)
    \item superresolution imaging
    \item object/entity recognition, localization, and tracking
    \item microstructure representation learning 
\end{itemize}

% Some of the common imaging techniques are SEM, SPM (STM/AFM), STEM, and TEM.
% A scanning electron microscope (SEM) provides images of a sample by scanning the surface with a focused beam of electrons.
% Scanning TEM (STEM) high an atomic number-contrast (Z-contrast) imaging where one can distinguish the elements directly from the contrast with high spatial resolution.
% A scanning probe microscope scan a tip over material's surface and detects electronic feedback signals in constant interaction mode or constant height modes.

Often these tasks generalize across many different imaging modalities, spanning optical microscopy (OM), scanning electron microscopy (SEM) techniques, scanned probe microscopy (SPM, as in scanning tunneling microscopy (STM) or atomic force microscopy (AFM), and transmission electron microscopy (TEM) variants, including scanning transmission electron microscopy (STEM).

The images obtained with these techniques range from capturing local atomic to mesoscale structures (microstructure), the distribution and type of defects and their dynamics which are critically linked to the functionality and performance of the materials.
Atomic-scale imaging has become widespread and near-routine over the past few decades  due to aberration corrected STEM \cite{Varela2005}.
Increasingly, collection of large image datasets is presenting an analysis bottleneck in the materials characterization pipeline, and the immediate need for automated image analysis becomes important. 
Non-DL image analysis methods have driven tremendous progress in quantitative microscopy, but often image processing pipelines are brittle and require too much manual identification of image features to be broadly applicable. Thus, DL is currently the most promising solution for high performance, high throughput automated analysis of image datasets.
For a good overview of applications in microstructure characterization specifically, see~\cite{Holm2020overview}.

\subsubsection{Databases and software libraries}

Image datasets for materials can come from either experiments or simulations. Software libraries mentioned above can be used to generate images such as STM/STEM. Images can also be obtained from the literature. A few common examples for image datasets is shown below in Table~\ref{tab:image-software}. Recently, there has been a rapid development in the field of image learning tasks for materials leading to several useful packages. We list some of them in Table~\ref{tab:image-software}.
%ZeroCostDL4Mic: Distributed and hosted computing for training and annotation~\cite{vonChamier2021}.%
\begin{table}[hbt!]
\caption{Databases and software packages for applying DL methods for image applications.}\label{tab:image-software}%
\begin{minipage}{174pt}

\begin{tabular}{@{}llll@{}}

\toprule
   & Databases  & \\
\midrule
DB Name   & Link  & Ref\\
\midrule
JARVIS-STM   &  \url{https://jarvis.nist.gov/jarvisstm}   & \cite{choudhary2021computational}  \\
atomagined   &  \url{https://github.com/MaterialEyes/atomagined}   & \cite{ophus2017fast}  \\ 
deep damage & \url{https://git.rwth-aachen.de/Sandra.Korte.Kerzel/DeepDamage} & \cite{kusche2019large} \\
NanoSEM & \url{https://doi.org/10.1038/sdata.2018.172} & \cite{aversa2018first} \\ 
UHCSDB & \url{http://hdl.handle.net/11256/940} & \cite{decost_hecht_francis_webler_picard_holm_2017} \\
UHCS micro. DB & \url{http://hdl.handle.net/11256/964} & \cite{decost_lei_francis_holm_2019} \\
SmBFO & \url{https://drive.google.com/} & \cite{ziatdinov2020causal} \\
Diffranet & \url{https://github.com/arturluis/diffranet} & \cite{deepfreak2019} \\
Peregrine v2021-03 & \url{https://doi.org/10.13139/ORNLNCCS/1779073} & \cite{osti_1779073} \\
Warwick electron\\microscopy data & \url{https://github.com/Jeffrey-Ede/datasets/wiki} & \cite{Ede2020db} \\

Powder bed\\anamoly  & \url{https://www.osti.gov/biblio/1779073} & \cite{osti_1779073} \\
% X: https://drive.google.com/uc?id=1PwaddcnZoXr_o2K_RTsVM1Uky8ov7yNT
%Y: https://git.rwth-aachen.de/Sandra.Korte.Kerzel/DeepDamage/
%######################################################################%
\midrule
   & Software packages  & \\
\midrule
Package Name   & Link  & Ref\\
\midrule

PyCroscopy   &  \url{https://github.com/pycroscopy/pycroscopy}   & \cite{somnath2019usid}  \\
Prismatic   &  \url{https://github.com/prism-em/prismatic}   & \cite{ophus2017fast}  \\
AtomVision   &  \url{https://github.com/usnistgov/atomvision}   & \cite{choudhary2021computational}  \\
py4DSTEM   &  \url{https://github.com/py4dstem/py4DSTEM}   & \cite{savitzky2020py4dstem}  \\
abTEM   &  \url{https://github.com/jacobjma/abTEM}   & \cite{madsen2021abtem}  \\
QSTEM   &  \url{https://github.com/QSTEM/QSTEM}   & \cite{koch2002determination}  \\
MuSTEM   &  \url{https://github.com/HamishGBrown/MuSTEM}   & \cite{allen2015modelling}  \\
MuSTEM   &  \url{https://github.com/HamishGBrown/MuSTEM}   & \cite{allen2015modelling}  \\
AICrystallographer   &  \url{https://github.com/pycroscopy/AICrystallographer}   & \cite{maxim_jesse_sumpter_kalinin_dyck_2020}  \\
AtomAI   &  \url{https://github.com/pycroscopy/atomai}   & \cite{maxim_jesse_sumpter_kalinin_dyck_2020}  \\
NionSwift   &  \url{https://github.com/nion-software/nionswift}   & \cite{meyer_dellby_hachtel_lovejoy_mittelberger_krivanek_2019}  \\
EENCM   &  \url{https://github.com/ceright1/Prediction-material-property}   & \cite{kim_tiong_kim_han_2021}  \\
DefectSegNet   &  \url{https://github.com/rajatsainju/DefectSegNet}   & \cite{roberts2019deep}  \\
AMPIS & \url{https://github.com/rccohn/AMPIS} & \cite{Cohn2021} \\
partial-STEM & \url{https://github.com/Jeffrey-Ede/partial-STEM/tree/1.0.0} & \cite{Ede2020} \\
ZeroCostDL4Mic & \url{https://github.com/HenriquesLab/ZeroCostDL4Mic} & \cite{von2020zerocostdl4mic} \\
EBSD-indexing & \url{https://github.com/NU-CUCIS/EBSD-indexing} & \cite{Jha2018} \\
PADNet-XRD & \url{https://github.com/NU-CUCIS/PADNet-XRD} & \cite{jha2019peak} \\
DKACNN & \url{https://github.com/NU-CUCIS/DKACNN} & \cite{yang2019deep} \\
PlasticityDL & \url{https://github.com/NU-CUCIS/PlasticityDL} & \cite{yang2020learning}\\
HomogenizationDL & \url{https://github.com/NU-CUCIS/HomogenizationDL} & \cite{yang2018deep}\\
LocalizationDL & \url{https://github.com/NU-CUCIS/LocalizationDL} & \cite{yang2019establishing}\\
MDGAN & \url{https://github.com/NU-CUCIS/MDGAN} & \cite{yang2018microstructural} \\
MDN-GAN & \url{https://github.com/NU-CUCIS/MDN-GAN} & \cite{yang2021general}\\

\botrule
\end{tabular}
\end{minipage}
\end{table}

% The deep learning approach allows researchers to train models on available labeled images (training set) which then can be used to predict accurate 1) image-level and/or 2) pixel level classification of new data. Some of the common software packages used for image classification/recognition tasks are: TorchVision, and OpenCV, Scikit-Image.

\subsubsection{Applications}
DL for images can be used to automatically extract information from images or transform images into a more useful state. The benefits of automated image analysis include higher throughput, better consistency of measurements compared to manual analysis, and even the ability to measure signals in images that humans cannot detect. The benefits of altering images include image super-resolution, denoising, inferring 3D structure from 2D images, and more. Examples of the applications of each task are summarized below.

%% image classification and regression
\paragraph{Image classification and regression}
Classification and regression are the processes of predicting one or more values associated with an image. In the context of DL the only difference between the two methods is that the outputs of classification are discrete while the outputs of regression models are continuous. The same network architecture may be used for both classification and regression by choosing the appropriate activation function (i.e., linear for regression or Softmax for classification) for the output of the network. Due to its simplicity image classification is one of the most established DL techniques available in the materials science literature. Nonetheless, this technique remains an area of active research.

Modarres et al. applied DL with transfer learning to automatically classify SEM images of different material systems \cite{modarres2017neural}. They demonstrated how a single approach can be used to identify a wide variety of features and material systems such as particles, fibers, Microelectromechanical systems (MEMS) devices, and more. The model achieved 90 \% accuracy on a test set. Misclassifications resulted from images that contained objects from multiple different classes, which is an inherent limitation of single-class classification. More advanced techniques like the ones described in subsequent sections can be applied to avoid these limitations.   Additionally, they developed a system to deploy the trained model at scale to process thousands of images in parallel. This approach is essential for large scale, high-throughput experiments or industrial applications of classification. ImageNet-based deep transfer learning has also been successfully applied for crack detection in macroscale materials images \cite{gopalakrishnan2017deep,gopalakrishnan2018crack}, as well as for property prediction on small, noisy, and heterogeneous industrial datasets \cite{yang2019data,yang2021heterogeneous}.  

DL has also been applied to characterize the symmetries of simulated measurements of samples. In ref \cite{Ziletti2018}, Ziletti et al. obtained a large database of perfect crystal structures, introduced defects into the perfect lattices, and simulated diffraction patterns for each structure. DL models were trained to identify the space group of each diffraction patterns. The model achieved high classification performance, even on crystals with significant numbers of defects, surpassing the performance of conventional algorithms for detecting symmetries from diffraction patterns. 

DL has also been applied to classify symmetries in simulated STM measurements of 2D material systems \cite{choudhary2021computational}.  DFT was used to generate simulated STM images for a variety of material systems. A convolutional neural network was trained to identify which of the five 2D Bravais lattices each material belonged to using the simulated STM image as input. The model achieved an average F1 score of around 0.9 for each lattice type. 

DL has also been used to improve the analysis of electron backscatter diffraction (EBSD) data, with Liu et al. \cite{liu2016materials} presenting one of the first DL-based solution for EBSD indexing capable of taking an EBSD image as input and predicting the three Euler angles representing the orientation that would have led to the given EBSD pattern. However, they considered the three Euler angles to be independent of each other, creating separate CNNs for each angle, although the three angles should really be considered together. Jha et al. \cite{Jha2018} built upon that work to train a single DL model to predict the three Euler angles in simulated EBSD patterns of polycrystalline Ni while directly minimizing the  misorientation angle between the true and predicted orientations. When tested on experimental EBSD patterns, the model achieved 16 \% lower disorientation error than dictionary based indexing. Similarly, Kaufman et al. trained a CNN to predict the corresponding space group for a given diffraction pattern \cite{Kaufmann2020PhaseID}. This enables EBSD to be used for phase identification in samples where the existing phases are unknown, providing a faster or more cost effective method of characterizing than X-ray or neutron diffraction. The results from these studies demonstrate the promise of applying DL to improve the performance and utility of EBSD experiments. 

Recently, DL has also been to learn crystal plasticity using images of strain profiles as input \cite{yang2019deep,yang2020learning}. The work in \cite{yang2019deep} used domain knowledge integration in the form of two-point auto-correlation to enhance the predictive accuracy, while \cite{yang2020learning} applied residual learning to learn crystal plasticity at nanoscale. It used strain profiles of materials of varying sample widths ranging from 2 $\mu m$ down to 62.5 $nm$ obtained from discrete dislocation dynamics to build a deep residual network capable of identifying prior deformation history of the sample as low, medium, or high. Compared to correlation function based method (68.24 \% accuracy), the DL model was found to be significantly more accurate (92.48 \%), and also capable of predicting stress-strain curves of test samples. This work also used saliency maps to try to interpret the developed DL model.

\paragraph{Pixelwise learning}
DL can also be applied to generate one or more predictions for every pixel in an image. This can provide more detailed information about the size, position, orientation, and morphology of features of interest in images.  Thus, pixelwise learning has been a significant area of focus with many recent studies appearing in materials science literature.

Azimi et al. applied an ensemble of fully convolutional neural networks to segment martensite, tempered martensite, bainite, and pearlite in SEM images of carbon steels.  Their model achieved 94 \% accuracy, demonstrating a significant improvement over previous efforts to automate the segmentation of different phases in SEM images. Decost, Francis, and Holm applied PixelNet to segment microstructural constituents in the UltraHigh Carbon Steel Database \cite{decost_hecht_francis_webler_picard_holm_2017, decost_lei_francis_holm_2019}. In contrast to fully convolutional neural networks, which encode and decode visual signals using a series of convolution layers, PixelNet constructs ``hypercolumns," or concatenations of feature representations corresponding to each pixel at different layers in a neural network. The hypercolumns are treated as individual feature vectors, which can then be classified using any typical classification approach, like a multi-layer perceptron.  This approach achieved phase segmentation precision and recall scores of 86.5 \% and 86.5 \%, respectively. Additionally, this approach was used to segment spheroidite particles in the matrix, achieving precision and recall scores of 91.1 \% and 91.1 \%, respectively. 

Pixelwise DL has also been applied to automatically segment dislocations in Ni superalloys \cite{Holm2020overview}. Dislocations are visually similar to $\gamma - \gamma^\prime$ and dislocation in Ni superalloys. With limited training data, a single segmentation model was unable to distinguish between these features. To overcome this, a second model was trained to generate a coarse mask corresponding to the deformed region in the material. Overlaying this mask with predictions from the first model selects the dislocations, enabling them to be distinguished from $\gamma - \gamma^\prime$ interfaces.

Stan, Thompson, and Voorhees applied Pixelwise DL to characterize dendritic growth from serial sectioning and synchrotron computed tomography data 
\cite{stan2020optimizing}.  Both of these techniques generate large amounts of data, making manual analysis impractical.  Conventional image processing approaches, utilizing thresholding, edge detectors, or other hand-crafted filters, are not able to deal with noise, contrast gradients, and other artifacts that are present in the data. Despite having a small training set of labeled images, SegNet was able to automatically segment these images with much higher performance.

\paragraph{Object/entity recognition, localization, and tracking}
Object detection or localization is needed when individual instances of recognized objects in a given image need to be distinguished from each other. In cases where instances do not overlap each other by a significant amount, individual instances can be resolved through post processing of semantic segmentation outputs. This technique has been applied extensively to the detection of individual atoms and defects in microstructural images.

Madsen et al. applied pixelwise DL to detect atoms in simulated atomic-resolution TEM images of graphene \cite{madsen2018deep}. A neural network was trained to detect the presence of each atom as well as predict its column height. Pixel-wise results are used as seeds for watershed segmentation to achieve instance-level detection. Analysis of the arrangement of the atoms led to autonomous characterization of defects in the the lattice structure of the material. Interestingly, despite being trained only on simulations, the model successfully detected atomic positions in experimental images.

Maksov et al. demonstrated atomistic defect recognition and tracking across sequences of atomic-resolution STEM images of WS$_2$ \cite{maksov2019deep}. The lattice structure and defects existing in the first frame were characterized through a physics-based approach utilizing Fourier transforms. The positions of atoms and defects in the first frame were used to train a segmentation model. Despite only using the first frame for training, the model successfully identified and tracked defects in the subsequent frames for each sequence, even when the lattice underwent significant deformation. Similarly, Yang et al.~\cite{yang2021deep} used U-net architecture (as shown in Fig. 4) to detect vacancies and dopants in WSe$_2$ in STEM images with model accuracy up to 98 \%. They classified the possible atomic sites based on experimental observations into five different types: tungsten, vanadium substituting for tungsten, selenium with no vacancy, mono-vacancy of selenium, and di-vacancy of selenium.

Roberts et al. developed DefectSegNet to automatically identify defects in transmission and STEM images of steel including dislocations, precipitates, and voids \cite{roberts2019deep}. They provide detailed information on the design, training, and evaluation of the model. They also compare measurements generated from the model to manual measurements performed by several different human experts, demonstrating that the measurements generated by DL are quantitatively more accurate and consistent. 

Kusche et al. applied DL to localize defects in panoramic SEM images of dual-phase steel \cite{kusche2019large}. Manual thresholding was applied to identify dark defects against the brighter matrix. Regions containing defects were classified via two neural networks. The first neural network distinguished between inclusions and ductile damage in the material. The second classified the type of ductile damage (i.e., notching, martensite cracking, etc.) Each defects was also segmented via watershed algorithm to obtain detailed information on its size, position, and morphology.
\begin{figure}
    \centering
    \includegraphics[width=1.0\textwidth]{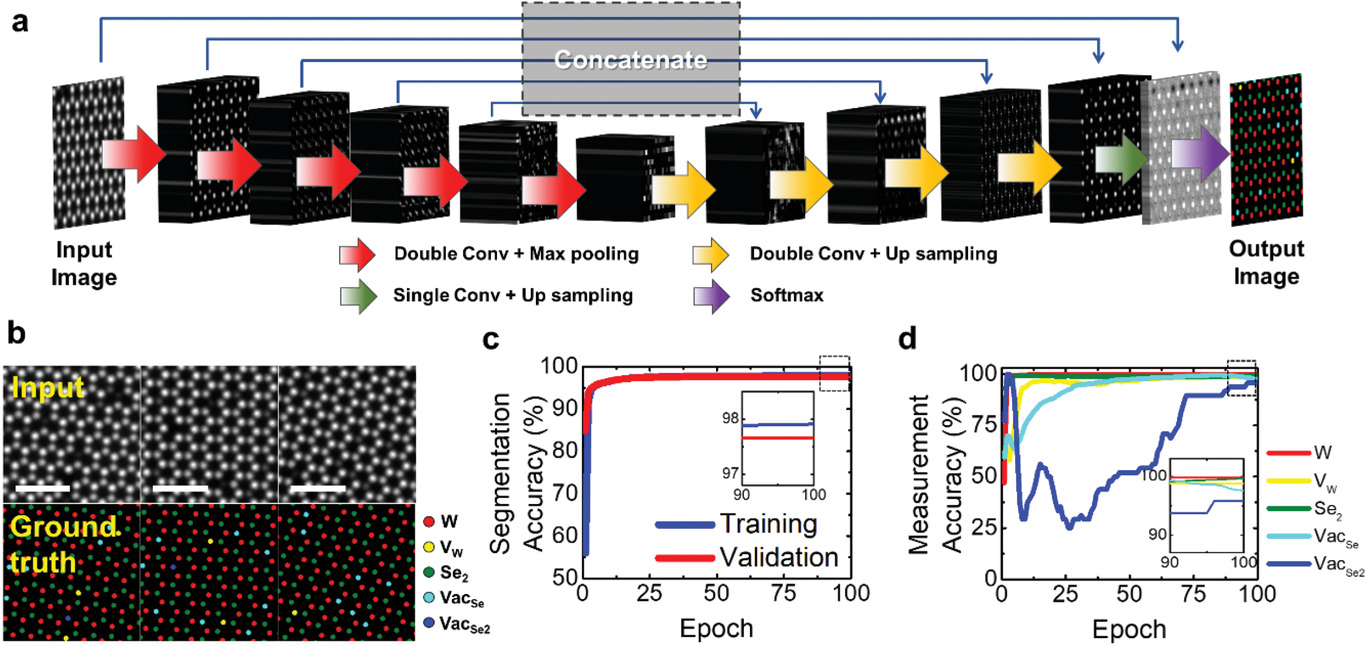}
    \caption{Deep learning-based algorithm for atomic site classification. a) Deep neural networks U-Net model constructed for quantification analysis of  annular dark-field in the scanning transmission electron microscope (ADF-STEM) image of V-WSe$_2$. b) Examples of training dataset for deep learning of atom segmentation model for five different species. c) Pixel-level accuracy of the atom segmentation model as a function of training epoch. d) Measurement accuracy of the segmentation model compared with human-based measurements. Scale bars are 1 nm [Reprinted according to the terms of the CC-BY license ref. \cite{yang2021deep}].}
\end{figure}

Applying DL to localize defects and atomic structures is a popular area in materials science research. Thus, several other recent studies on these applications can be found in the literature \cite{vlcek2019learning, ziatdinov2017learning, Ovchinnikov2020,LiW2018}.

In the above examples pixelwise DL, or classification models are combined with image analysis to distinguish individual instances of detected objects. However, when there are several adjacent objects of the same class that touch or overlap each other in the image, this approach will falsely detect them to be a single, larger object. In this case, DL models designed for detection or instance segmentation can be used to resolve overlapping instances. In one such study, Cohn and Holm applied DL for instance level segmentation of individual particles and satellites in dense powder images \cite{Cohn2021}.  Segmenting each particle allows for computer vision to generate detailed size and morphology information which can be used to supplement experimental powder characterization for additive manufacturing. Additionally, overlaying the powder and satellite masks yielded the first method for quantifying the satellite content of powder samples, which cannot be measured experimentally.

\paragraph{Superresolution imaging and auto-tuning experimental parameters}

The studies listed so far focus on automating the analysis of existing data after it has been collected experimentally. However, DL can also be applied during experiments to improve the quality of the data itself. This can reduce time for data collection or improve the amount of information captured in each image. Super-resolution and other DL techniques can also be applied in-situ to autonomously adjust experimental parameters. 

Recording high-resolution electron microscope images often requires large dwell times, limiting the throughput of microscopy experiments. Additionally, during imaging, interactions between the electron beam and a microscopy sample can result in undesirable effects including charging of non-conductive samples and damaging of sensitive samples. Thus, there is interest in using DL to artificially increase the resolution of images without introducing these artifacts. One method of interest is applying generative adversarial networks (GANs) for this application.

De Haan et al. recorded SEM images of the same regions of interest in carbon samples containing gold nanoparticles at two different resolutions \cite{de2019resolution}. Low-resolution images recorded at were used as inputs to a GAN.  The corresponding images with twice the resolution were used as the ground truth. After training the GAN reduced the number of undetected gaps between nanoparticles from 13.9 \% to 3.7 \%, indicating that super-resolution was successful. Thus, applying DL led to a four-fold reduction of the interaction time between the electron beam and the sample.

Ede and Beanland collected a dataset of STEM images of different samples \cite{Ede2020}. Images were subsampled with spiral and `jittered' grid masks to obtain partial images with resolutions reduced by a factor up to 100.  A GAN was trained to reconstruct full images from their corresponding partial images. The results indicated that despite a significant reduction in the sampling area, this approach successfully reconstructed high resolution images with relatively small errors. 

DL has also been applied to automated tip conditioning for SPM experiments. Rashidi and Wolkow trained a model to detect artifacts in SPM measurements resulting from a degredation in tip quality \cite{rashidi2018autonomous}. Using an ensemble of convolutional neural networks resulted in 99 \% accuracy. After detecting that a tip has degraded, the SPM was configured to automatically recondition the tip \emph{in-situ} until the network indicated that the atomic sharpness of the tip has been restored. Monitoring and reconditioning the tip is the most time and labor intensive part of conducting SPM experiments. Thus, automating this process through DL can increase the throughput and decrease the cost of collecting data through SPM.

In addition to materials characterization, DL can be applied to autonomously adjust parameters during manufacturing. Scime et al. mounted a camera to multiple 3D printers  \cite{Scime2020}. Images of the build plate were recorded throughout the printing process. A dynamic segmentation convolutional neural network was trained to recognize defects such as recoater streaking, incomplete spreading, spatter, porosity, and others. The trained model achieved high performance and was transferable to multiple printers from three different methods of additive manufacturing. This work is the first step to enabling smart additive manufacturing machines that can correct defects and adjust parameters during printing.

There is also growing interest in establishing instruments and laboratories for autonomous experimentation. Eppel et al. trained multiple models to detect chemicals, materials, and transparent vessels in a chemistry lab setting \cite{eppel2020}.  This study provides a rigorous analysis of several different approaches for scene understanding. Models were trained to characterize laboratory scenes with different methods including semantic segmentation and instance segmentation, both with and without overlapping instances. The models successfully detected individual vessels and materials in a variety of settings. Finer-grained understanding of the contents of vessels, such as segmentation of individual phases in multi-phase systems, was limited, outlining the path for future work in this area. The results is an important step towards the development of automated experimentation for laboratory scale experiments. 

 % TODO: setting diffraction conditions and other hyperparameters. LeBeau group maybe?

\paragraph{Microstructure representation learning}

% Image-based DL models are very large and may have up to 200 million parameters. Thus, training models can require enormous labeled datasets. For example, ImageNet \cite{Russakovsky2015} currently contains over 14 million labeled images for classification.  Obtaining datasets for materials applications, especially in smaller academic studies, is clearly infeasible. Thus, many studies applying image based DL models to materials science application leverage a technique called transfer learning. After training models on one dataset, such as ImageNet, models can be adapted to different applications, such as  analysis of images in materials science, with significantly smaller datasets in the target domain.

% Transfer learning is believed to be especially suitable for images because images containing different objects and scenes share lower-level image features like edges, corners, and `blobs.' After learning representations that capture these lower-level features, neural networks can be adapted to new applications without needing as much training data. Transfer learning is an established technique in DL and appears in many of the studies cited in this review. Despite this, DL models are viewed as `black-boxes.'  Interpretations of the internal representations of image based DL models remains an open question in both DL and materials science research.

Materials microstructure is often represented in the form of multi-phase high-dimensional 2D/3D images and thus can readily leverage image-based DL methods to learn robust, low-dimensional microstructure representations, which can subsequently be used for building predictive and generative models to learn forward and inverse structure-property linkages, which are typically studied across different length scales (multi-scale modeling). In this context, homogenization and localization refer to transfer of information from lower length scales to higher length scales and vice-versa. DL using customized CNNs has been used both for homogenization, i.e., predicting the macroscale property of a material given its microstructure information \cite{yang2018deep,cecen2018material,yang2019deep}, as well as for localization, i.e., predicting the strain distribution across a given microstructure for a loading condition \cite{yang2019establishing}.

Transfer learning has also been widely used for analyzing materials microstructure images, and methods for improving the use of transfer learning to materials science applications is still an area of active research.  Goetz et al. investigated the use of unsupervised domain adaptation as an alternative to simply fine-tuning a pre-trained model \cite{goetz2021}. In this technique a model is first trained on a labeled dataset in the source domain. Next, a discriminator model is used to train the model to generate features that are domain-agnostic. Comapared to simple fine-tuning, unsupervised domain adaptation improved the performance of classification and segmentation neural networks on materials science datasets. However, it was determined that the highest performance was achieved when the source domain was more visually similar to the target (for example, using a different set of microstructural images instead of ImageNet.) This highlights the utility of establishing large, publicly available datasets of annotated images in materials science.

Kitaraha and Holm used the output of an intermediate layer of a pre-trained convolutional neural network as a feature representation for images of steel surface defects and Inconnel fracture surfaces \cite{Kitahara2018}. Images were classified by defect type or fracture surface orientation, respectively, using unsupervised DL. Even though no labeled data was used for training the neural network or the unsupervised classifier, the model found natural decision boundaries that achieved classification performance of 98 \% and 88 \% for the defect classes and fracture surface orientations, respectively.  Visualization of the representations through principal component analysis (PCA) and t-distributed stochastic neighborhood embedding (t-SNE) provided qualitative insights into the representations. Though detailed physical interpretation of the representations is still a distant goal, this study provides tools for investigating patterns in visual signals contained in image-based datasets in materials science.

Larmuseau et al. investigated the use of triplet networks to obtain consistent representations for visually similar images of materials  \cite{Larmuseau2020}. Triplet networks are trained with three images at a time. The first image, the reference, is classified by the network. The second image, called the positive, is another image with the same class label. The last image, called the negative, is an image from a separate class. During training the loss function includes errors in prediction of the class of the reference image, the difference in representations of the reference and positive images, and the similarity in representations of the reference and negative images. This process allows the network to learn representations that are consistent for images in the same class while distinguishing images from different classes. The triple network outperformed an ordinary convolutional neural network trained for image classification on the same dataset.

% Microstruture representation learning using Siamese networks \cite{Sardeshmukh2020} %% note: Looks extremely relevant but CMU apparently doesn't have access to this paper. 

In addition to investigating representations used to analyze existing images, DL can be applied to generate synthetic images of materials systems. Generative Adversarial Networks (GANs) are currently the predominant method for synthetic microstructure generation. GANs consist of a generator, which create a synthetic microstructure image, and a discriminator, which attempts to predict if a given input image is real or synthetic. With careful application, GANs can be used as a powerful tool for microstructure representation learning and design. 

Yang and Li et al. \cite{li2018deep,yang2018microstructural} developed a GAN-based model for learning a low-dimensional embedding of microstructures, which could then be easily sampled and used with the generator of the GAN model to generate realistic, statistically similar microstructure images, thus enabling microstructural materials design. The model was able to capture complex, non-linear microstructure characteristics and learn the mapping between the latent design variables and microstructures. In order to close the loop, the method was combined with a Bayesian optimization approach to design microstructures with optimal optical absorption performance. The discovered microstructures were found to have up to 17 \% better property than randomly sampled microstructures. The unique architecture of their GAN model also facilitated generator scalability to generate arbitrary sized microstructure images and discriminator transferability to build structure-property prediction models. Yang et al. \cite{yang2021general} recently combined GANs with MDNs (mixture density networks) to enable inverse modeling in microstructural materials design, i.e., generate the microstructure for a given desired property.  

Hsu et al. constructed a GAN to generate 3D synthetic solid oxide fuel cell microstructures~\cite{Hsu2020}. These microstructures were compared to other synthetic microstructures generated by DREAM.3D as well as experimentally observed microstructures measured via sectioning and imaging with PFIB-SEM.  Synthetic microstructures generated from the GAN were observed to qualitatively show better agreement to the experimental microstructures than the DREAM.3D microstructures, as evidenced by the more realistic phase connectivity and lower amount of agglomeration of solid phases. Additionally, a statistical analysis of various features such as volume fraction, particle size, and several other quantities demonstrated that the GAN microstructures were quantitatively more similar to the real microstructures than the DREAM.3D microstructures.

In a similar study, Chun et al. generated synthetic microstructures of high energy materials using a GAN \cite{Chun2020}. Once again, a synthetic microstructure generated via GAN showed better qualitative visual similarity to an experimentally observed microstructure compared to a synthetic microstructure generated via a transfer learning approach, with sharper phase boundaries and fewer computational artifacts. Additionally, a statistical analysis of the void size, aspect ratio, and orientation distributions indicated that the GAN produced microstructures that were quantitatively more similar to real materials.

Applications of DL to microstructure representation learning can help researchers improve the performance of predictive models used for the applications listed above. Additionally, using generative models can generate more realistic simulated microstructures. This can help researchers develop more accurate models for predicting material properties and performance without needing to actually synthesize and process these materials, significantly increasing the throughput of materials selection and screening experiments.

% \paragraph{Unfiled references}

% infer 3D rotation distortions of atoms of BO6 polyhedron \cite{he2015towards}.

% semantic segmentation \(\rightarrow\) parameterizing kinetic model for precipitation reaction~\cite{Kautz2020}.

%%% ADDED BY RYAN ON 2021-10-20
\paragraph{Mesoscale modeling applications}
% TODO: phase field and deep learning
In addition to image-based characterization, deep learning methods are increasingly used in mesoscale modeling.
Dai et al.~\cite{Dai2021} trained a GNN successfully trained to predict magnetostriction in a wide range of synthetic polycrystalline systems with around 10 \% prediction error.
The microstructure is represented by a graph where each node correspond to a single grain, and the edges between nodes indicate an interface between neighboring grains.
Five node features (3 Euler angles, volume, and number of neighbors) were associated with each grain. The GNN was able to outperform other machine learning approaches for property prediction of polycrystalline materials by accounting for interactions between neighboring grains. 
% Additionally, the authors note that the trained GNN is able to generate predictions on new microstructures in a fraction of a second. In contrast, the physics-based simulations used to generate the data required 5 hours on an advanced supercomputer to generate a single data point.  

Similarly, Cohn and Holm present preliminary work applying GNNs to predict the occurrence of abnormal grain growth (AGG)  in Monte Carlo simulations of microstructure evolution~\cite{Cohn2021gnn}. AGG appears to be stochastic, making it notoriously difficult to predict, control, and even observe experimentally in some materials. AGG has been reproduced in Monte Carlo simulations of material systems, but model that can to predict which initial microstructures will undergo AGG has not been established before. A dataset of Monte Carlo simulations was created using SPPARKS \cite{SPPARKS,Plimpton2009}. A microstructure GNN was trained to predict AGG in individual simulations, with 75 \% classification accuracy. In comparison, an image-based only achieved 60 \% accuracy. The GNN also provided physical insight to understanding AGG and indicated that only 2 neighborhood shells are needed to achieve the maximum performance achieved in the study. These early results motivate additional work on applying GNNs to predict the occurence in both simulated and real materials during processing.

%% Dai et al dataset: https://github.com/mehmetfdemirel/PolycrystalGraph
%% Dai et al code: https://github.com/mehmetfdemirel/PolycrystalGraph

%% Cohn and Holm dataset and code: Not yet ready for release, but will be released when completed study is ready for publication.
%%%%%

\subsection{Natural language processing}\label{sec:nlp}

Most of existing knowledge in the materials domain is currently unavailable as structured information and only exists as unstructured text, tables or images in various publications. There exists a great opportunity to use natural language processing (NLP) techniques to convert text to structured data or to directly learn and make inferences from text information. However, as a relatively new field within materials science, many challenges remain unsolved in this domain, such as how to resolve dependencies between words and phrases across multiple sentences and paragraphs.

\subsubsection{Data sets for NLP}

Data sets relevant to natural language processing include peer-reviewed journal articles, articles published on preprint servers such as arXiv or ChemRxiv, patents, and online material such as Wikipedia. Unfortunately, being able to access or use most such data sets remains difficult. Peer-reviewed journal articles are typically subject to copyright restrictions and thus difficult to obtain, especially in the large numbers required for machine learning. Many publishers now offer text and data mining (TDM) agreements that can be signed online, and which allow at least a limited, restricted amount of work to be performed. However, gaining access to the full text of a large number of publications still typically requires strict and dedicated agreements with each publisher. The major advantage of working with publishers is that they have often already converted the articles from a document format such as PDF into an easy-to-parse format such as HyperText Markup Language (HTML). In contrast, articles on preprint servers and patents are typically available with fewer restrictions, but are typically available only as PDF files. Currently, it remains difficult to properly parse text from PDF files in a reliable manner, even when the text is embedded in the PDF. Therefore, new tools that can easily and automatically convert such content into well-structured HTML format with few residual errors would likely have a major impact on the field. Finally, online sources of information such as Wikipedia can serve as another type of data source, however often such online sources are more difficult to verify in terms of accuracy and also do not contain as much domain-specific information as the research literature.

\subsubsection{Software libraries for NLP}

Applying NLP to a raw data set involves multiple steps, including retrieving the data, various forms of ``pre-processing" (sentence and word tokenization, word stemming and lemmatization, featurization such as word vectors or part of speech tagging), and finally machine learning for information extraction (e.g., named entity recognition, entity relationship modeling, question and answer, or others). There exist multiple software libraries to aid in materials NLP, as described in Table~\ref{tab:nlp-sw}.
We note that although many of these steps can in theory be performed by general-purpose NLP libraries such as NLTK~\cite{xue_2010}, SpaCy~\cite{spacy2}, or AllenNLP~\cite{Gardner2017AllenNLP}, the specialized nature of chemistry and materials science text (including the presence of complex chemical formulas) often leads to errors. For example, researchers have developed specialized codes to perform pre-processing that better detect chemical formulas (and not split them into separate tokens or apply stemming/lemmatization to them) and scientific phrases and notation such as oxidation states or symbols for physical units. Similarly, chemistry-specific codes for extracting entities are better at extracting the names of chemical elements (e.g., recognizing that ``He" likely represents helium and not a male pronoun) and abbreviations for chemical formulas. Finally, word embeddings that convert words such as ``manganese" into numerical vectors for further data mining are more informative when trained specifically on materials science text versus more generic texts, even when the latter data sets are larger \cite{tshitoyan2019unsupervised}. Thus, domain-specific tools for NLP are required in nearly all aspects of the pipeline. The main exception is that the architecture of the specific neural network models used for information extraction (e.g., LSTM, BERT, or architectures used to generate word embeddings such as word2vec or GloVe) are typically not modified specifically for the materials domain. Thus, much of the materials and chemistry-centric work currently regards data retrieval and appropriate preprocessing. A longer discussion of this topic, with specific examples, can be found in refs. \cite{kononova2021opportunities,olivetti2020data}.

\begin{table}[hbt!]

% \begin{minipage}{174pt}
\begin{minipage}{0.9\textwidth}
\caption{Software packages for applying DL methods for natural language processing}\label{tab:nlp-sw}%
% \begin{tabular}{@{}llll@{}}
 \begin{tabular}{@{}lll@{}}
\toprule

Software name  & Link  & Ref\\
\midrule
Borges    & \url{https://github.com/CederGroupHub/Borges}   & \cite{he2020similarity}  \\

ChemDataExtractor  & \url{http://chemdataextractor.org}  & \cite{swain2016chemdataextractor}  \\

ChemicalTagger  & \url{https://github.com/BlueObelisk/chemicaltagger}  & \cite{hawizy2011chemicaltagger}  \\

ChemListem   & \url{https://bitbucket.org/rscapplications/chemlistem/}  & \cite{corbett2018chemlistem}  \\

ChemSpot   & \url{https://github.com/rockt/ChemSpot}  & \cite{rocktaschel2012chemspot}  \\

LBNLP    & \url{https://github.com/lbnlp/lbnlp} & \cite{weston2019named}  \\

mat2vec   & \url{https://github.com/materialsintelligence/mat2vec} & \cite{tshitoyan2019unsupervised}  \\

MaterialsParser     & \url{https://github.com/CederGroupHub/MaterialParser} & \cite{kononova2019text}  \\

OSCAR4    & \url{https://github.com/BlueObelisk/oscar4}  & \cite{jessop2011oscar4}  \\

Synthesis Project   & \url{https://www.synthesisproject.org} & \cite{kim2017materials}  \\

tmChem    & \url{https://www.ncbi.nlm.nih.gov/research/bionlp/Tools/tmchem/}  & \cite{leaman2015tmchem}  \\

\botrule
\end{tabular}
\end{minipage}
\end{table}

\begin{figure}
    \centering
    \includegraphics[trim={0 .1cm 0 1cm},clip,width=1\textwidth]{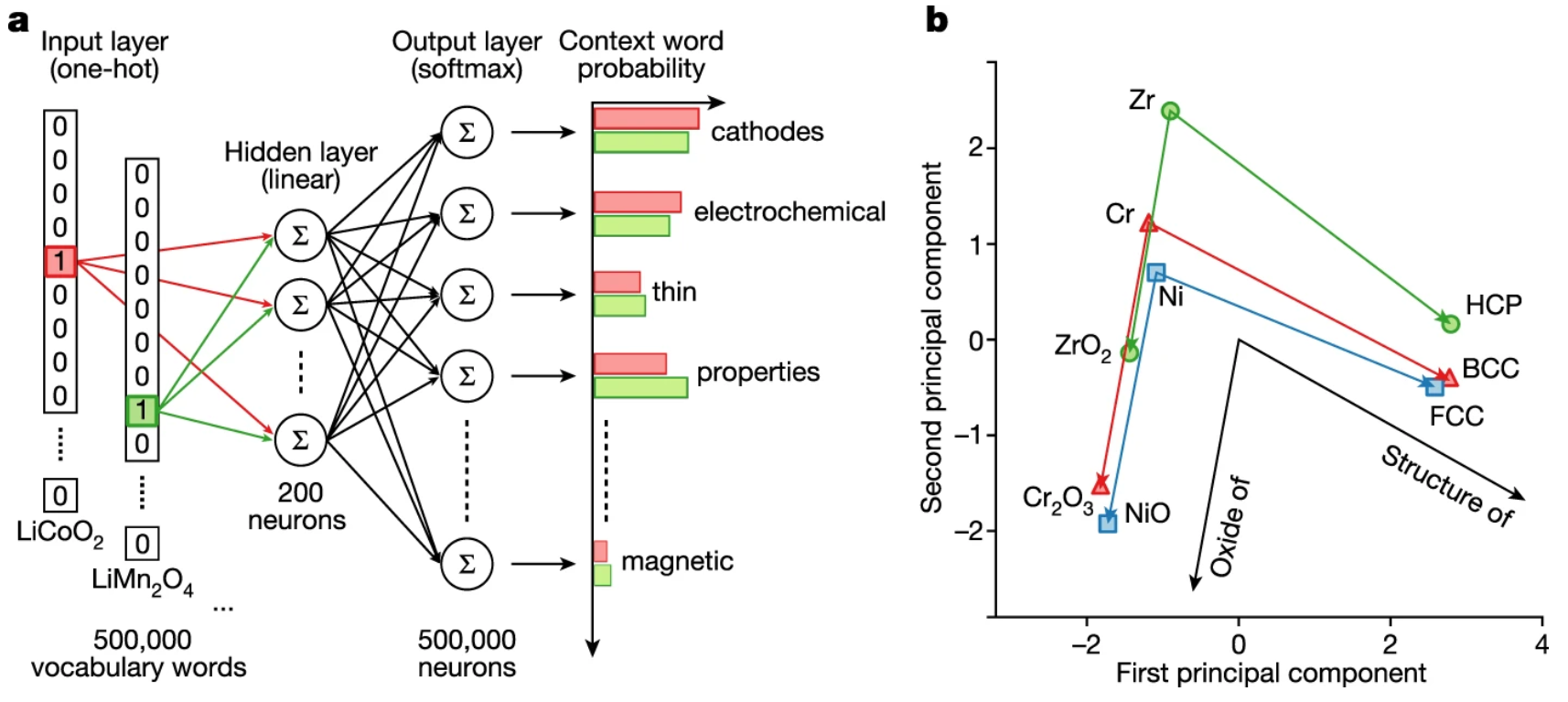}
    \caption{Left: network for training word embeddings for natural language processing application. A one-hot encoded vector at left represents each distinct word in the corpus; the role of a hidden layer is to predict the probability of neighboring words in the corpus. This network structure trains a relatively small hidden layer of 100 to 200 neurons to contain information on the context of words in the entire corpus, with the result that similar words end up with similar hidden layer weights (word embeddings). Such word embeddings can be used to transform textual words into numerical vectors useful for a variety of applications. Right: projection of word embeddings for various materials science words, as trained on a corpus scientific abstracts, into two dimensions using principle components analysis. Without any explicit training, the word embeddings naturally preserve relationships between chemical formulas, their common oxides, and their ground state structures. [Reprinted according to the terms of the CC-BY license ref. \cite{tshitoyan2019unsupervised}]}
\end{figure}

\subsubsection{Applications}
NLP methods for materials have been applied for information extraction and search (particularly as applied to synthesis prediction) as well as materials discovery. As the domain is rapidly growing, we suggest dedicated reviews on this topic by Olivetti et al. \cite{olivetti2020data} and Kononova et al. \cite{kononova2021opportunities} for more information.

One of the major uses of NLP methods is to extract data sets from text in published studies. Conventionally, such data sets required manual entry of data sets by researchers combing the literature, a laborious and time-consuming process. Recently, software tools such as ChemDataExtractor \cite{swain2016chemdataextractor} and other methods \cite{park2018text} based on more conventional machine learning and rule-based approaches have enabled automated or semi-automated extraction of data sets such as Curie and Néel magnetic phase transition temperatures \cite{court2018auto}, battery properties \cite{huang2020database}, UV-vis spectra \cite{beard2019comparative}, and surface and pore characteristics of metal organic frameworks \cite{tayfuroglu2019silico}. In the past few years, DL approaches such as LSTMs and transformer-based models have been employed to extract various categories of information \cite{weston2019named}, and in particular materials synthesis information \cite{vaucher2020automated,he2020similarity,kononova2019text} from text sources. Such data has been used to predict synthesis maps for titania nanotubes \cite{kim2017materials}, various binary and ternary oxides \cite{kim2017virtual}, and perovskites \cite{kim2020inorganic}. 

Databases based on natural language processing have also been used to train machine learning models to identify materials with useful functional properties, such as the recent discovery of the large magnetocaloric properties of HoBe$_2$ \cite{de2020machine}. Similarly, Cooper et al. \cite{cooper2019design} demonstrated a “design to device approach” for designing dye-sensitized solar sells that are co-sensitized with two dyes \cite{cooper2019design}. This study used automated text mining to compile a list of candidate dyes for the application along with measured properties such as maximum absorption wavelengths and extinction coefficients. The resulting list of 9431 dyes extracted from the literature were downselected to 309 candidates using a variety of criteria such as molecular structure and ability to absorb in the solar spectrum. These candidates were evaluated for suitable combinations for co-sensitization, yielding 33 dyes that were further downselected using density functional theory calculations and experimental constraints. The resulting 5 dyes were evaluated experimentally, both individually and in combinations, resulting in a combination of dyes that not only outperformed any of the individual dyes but demonstrated performance comparable to an existing standard material. This study demonstrates the possibility of using literature-based extraction to identify materials candidates for new applications from the vast body of published work, which may have never tested those materials for the desired application.

It is even possible that natural language processing can directly make materials predictions without the use of intermediary models. In a study reported by Tshitoyan et al. \cite{tshitoyan2019unsupervised} (as shown in Fig. 5), word embeddings (i.e., numerical vectors representing distinct words) trained on materials science literature could directly predict materials applications through a simple dot product between the trained embedding for a composition word (such as PbTe) and an application words (such as thermoelectrics). The researchers demonstrated that such an approach, if applied in the past using historical data, may have subsequently predicted many recently reported thermoelectric materials; they also presented a list of potentially interesting thermoelectric compositions using the known literature at the time. Since then, several of these predictions have since been tested either computationally \cite{yang2018low,wang2019ultralow,jong2020manifestation,yamamoto2020first,viennois2020anisotropic,haque2020effect} or experimentally\cite{yahyaoglu2021phase} as potential thermoelectrics. Recently, such approaches have also been applied to search for understudied areas of metallocene catalysis \cite{ho2020using}, although challenges still remain in such direct approaches to materials prediction.

\section{Uncertainty quantification}\label{sec:uncertainty}

Uncertainty quantification (UQ) is an essential step in the evaluation of the robustness of DL. Specifically, DL models have been criticized for lack of robustness, interpretability, and reliability and the addition of carefully quantified uncertainties would go a long way towards addressing such shortcomings. While most of the focus in the DL field currently goes into developing new algorithms or training networks to high accuracy, there is an increasing attention to UQ, as exemplified by the detailed review of Abdar et al. \cite{abdar2021review}. However, determining the uncertainty associated to DL predictions is still a challenging and far from a completely solved problem. 

The main drawback to estimating UQ when performing DL is the fact that most of the currently available UQ implementations do not work for arbitrary, off-the-shelf models, without retraining or redesigning. Bayesian NNs are the exception; however, they require significant modifications to the training procedure, are computationally expensive compared to non-Bayesian NNs and become increasingly inefficient the larger the data size gets. A significant fraction of the current research in DL UQ focuses exactly on such an issue: how to evaluate uncertainty without requiring computationally expensive re-training or DL code modifications. An example of such an effort is the work of Mi et al \cite{mi1910training}, where three scalable methods are explored, to evaluate the variance of output from trained NN, without requiring any amount of re-training. Another example is Teye, Azizpour and Smith’s exploration of the use of batch normalization as a way to approximate inference in Bayesian models \cite{teye2018bayesian}.  

Before reviewing the most common methods used to evaluate uncertainty in DL, let us briefly point out key reasons to add UQ to DL modeling. Reaching high accuracy when training DL models implicitly assumes the availability of a sufficiently large and diverse training dataset. Unfortunately, this rarely occurs in material discovery applications \cite{zhang2021leveraging}. ML/DL models are prone to perform poorly on extrapolation \cite{meredig2018can} . They also find extremely difficult to recognize ambiguous samples \cite{zhang2020mix}. In general, determining the amount of data necessary to train a DL to achieve the required accuracy is a challenging problem. Careful evaluation of the uncertainty associated with DL predictions would not only increase reliability in predicted results but would also provide guidance on estimating the needed training data set size as well as suggesting what new data should be added to reach the target accuracy (uncertainty-guided decision). Zhang, Kailkhura, and Han’s work emphasizes how including a UQ-motivated reject option into the DL model results in substantial improvements in the performance of the remaining material data \cite{zhang2021leveraging}. Such a reject option is associated to the detection of out-of-distribution samples, which is only possible through UQ analysis of the predicted results.

Two different uncertainty types are associated with each ML prediction: epistemic uncertainty and aleatory uncertainty. Epistemic uncertainty is related to insufficient training data in part of the input domain. As mentioned above, while DL are very effective at interpolation tasks, they cannot extrapolate, and, therefore, it’s vital to quantify the lack of accuracy due to localized, insufficient training data. The aleatory uncertainty, instead, is related to parameters not included in the model. It relates to the possibility of training on data that our DL perceives as very similar but that are associated to different outputs because of missing features in the model. Ideally, we would like UQ methodologies able to distinguish, and separately quantify, both types of uncertainties.

The most common approaches to evaluate uncertainty using DL are Dropout methods, Deep Ensemble methods, Quantile regression and Gaussian Processes.
Dropout methods are commonly used to avoid over-fitting. In this type of approach, network nodes are disabled randomly during training, resulting in evaluation of a different subset of the network at each training step. When a similar randomization procedure is applied to the prediction procedure as well, the methodology becomes Monte-Carlo dropout \cite{seoh2020qualitative}. Repeating such randomization multiple times produces a distribution over the outputs, from which mean and variance are determined for each prediction. Another example of using a dropout approach to approximate Bayesian inference in deep Gaussian processes is the work of Gal and Ghahramani \cite{gal2016dropout}.

Deep ensemble methodologies \cite{jain2020maximizing,ganaie2021ensemble,fort2019deep,lakshminarayanan2016simple} combine deep learning modelling with ensemble learning. Ensemble methods utilize multiple models and different random initializations to improve predictability. Because of the multiple predictions, statistical distributions of the outputs are generated. Combining such results into a Gaussian distribution, confidence intervals are obtained through variance evaluation. Such a multi-model strategy allows the evaluation of aleatory uncertainty when sufficient training data are provided. For areas without sufficient data, the predicted mean and variance will not be accurate, but the expectation is that a very large variance should be estimated, clearly indicating non-trustable predictions. Monte-Carlo Dropout and Deep Ensembles approaches can be combined to further improve confidence in the predicted outputs. 

Quantile regression can be utilized with DL \cite{moon2021learning}. In this approach, the loss function is used in a way that allows to predict for the chosen quantile a (between 0 and 1). A choice of $a = 0.5$ corresponds to evaluating the Mean Absolute Error (MAE) and predicting the median of the distribution. Predicting for two more quantile values (amin and amax) determines confidence intervals of width amax – amin. For instance, predicting for amin = 0.1 and amax = 0.8 produces confidence intervals covering 70 \% of the population. The largest drawback of using quantile to estimate prediction intervals is the need to run the model 3 times, one for each quantile needed. However, a recent implementation in TensorFlow allows to simultaneously obtain multiple quantiles in one run.

Lastly, Gaussian Processes (GP) can be used within a DL approach as well and have the side benefit of providing UQ information at no extra cost. Gaussian processes are a family of infinite-dimensional multivariate Gaussian distributions completely specified by a mean function and a flexible kernel function (prior distribution). By optimizing such functions to fit the training data, the posterior distribution is determined, which is later used to predict outputs for inputs not included in the training set. Because the prior is a Gaussian process, the posterior distribution is Gaussian as well \cite{rasmussen2003gaussian}, thus providing mean and variance information for each predicted data. However, in practice standard kernels under-perform \cite{hegde2018deep}. In 2016, Wilson et al. \cite{wilson2016deep} suggested to process inputs through a neural network prior to a Gaussian process model. This allowed to extract high-level patterns and features, however required careful design and optimization. In general, Deep Gaussian processes improve the performance of Gaussian processes by mapping the inputs through multiple Gaussian process `layers'. Several groups have followed this avenue and further perfected such an approach (\cite{hegde2018deep} and references within). A common drawback of Bayesian methods is a prohibitive computational cost if dealing with large datasets \cite{gal2016dropout}.

\section{Limitations and challenges}\label{sec:challenges}
 
Although DL methods have various fascinating opportunities for materials design, they have several limitations and there is much room to improve. Reliability and quality assessment of datasets used in DL tasks are challenging because there is either a lack of a ground truth data, or there are not enough metrics for a global comparison, or datasets using similar or identical set-ups may not be reproducible \cite{hegde2020reproducibility}. This poses an important challenge on relying upon DL based prediction. 

Material representations based on chemical formula alone by definition do not consider structure, which on the one hand makes them more amenable to work for new compounds for which structure information may not be available, but on the other hand makes it impossible for them to capture phenomena such as phase transitions.  Properties of materials depend sensitively on structure to the extent that their properties can be quite opposite depending on the atomic arrangement, like diamond (hard, wide-band-gap insulator) and graphite (soft, semi-metal). It is thus not a surprise that chemical formula based methods may not be adequate in some cases \cite{bartel2020critical}. 

Atomistic graph based predictions, though considered a full atomistic description, are tested on bulk materials only and not for defective systems or for multi-dimensional phases space exploration such as using genetic algorithms.  In general, this underscores that the input features must be predictive for the output labels and not be missing some key information. Although atomistic graph neural network models such as atomistic line graph neural network (ALIGNN) have achieved remarkable accuracy compared to previous atomistic based models, the model errors still need to be further brought down to reach something resembling deep-learning `chemical-accuracies.'

In terms of images and spectra, the experimental data are too noisy most of the time and require much manipulation before applying DL, while theory based simulated data work, but being noise-free do not capture realistic scenarios \cite{choudhary2021computational}. 

Uncertainty quantification for deep learning for materials science is important and yet only a few works have been done in this field. To alleviate the black box \cite{holm2019defense} nature of the DL methods, package such as GNNExplainer \cite{ying2019gnnexplainer} has been tried in the materials context. Such attempts at greater interpretability will be important moving forward to gain the trust of the materials community. 

While training-validation-test split strategies were primarily designed in DL for image classification tasks with a certain number of classes, the same for regression models in materials science may not be the best approach. This is because it is possible that that during the training the model is seeing a material very similar to the test set material and in reality it is difficult to generalize the model. Best practices need to be developed for data split, normalization and augmentation to avoid such issues \cite{meredig2018can}.

Finally, we note an important technological challenge is to make a ``closed-loop'' autonomous materials design and synthesis process \cite{roch2018chemos,szymanski2021toward} that can include both machine learning and experimental components in a ``self-driving laboratory''~\cite{macleod2020self}. For an overview of early proof of principle attempts see \cite{stach;m21}. For example, in an autonomous synthesis experiment the oxidation state of copper  (and therefore the oxide phase) was varied in a sample of copper oxide by automatically flowing more oxidizing or more reducing gas over the sample and monitoring the charge state of the copper using XANES.  An algorithmic decision policy was then used to automatically change the gas composition for a subsequent experiment based on the prior experiments, with no human in the loop, in such a way as to autonomously move towards a target copper oxidation state \cite{rakit;jacs20}. This is a simple proof of principle experiment that gives just a glimpse of what is possible moving forward.

% final thoughts

\section{Contributions}
The authors contributed equally to the search as well as analysis of the literature and writing of the manuscript.
\section{Competing interests}
The authors declare no competing interests.

\section{Acknowledgments}
Contributions from K.C. were supported by the financial assistance award 70NANB19H117 from the U.S. Department of Commerce, National Institute of Standards and Technology.
E.A.H. and R.C. (CMU) were supported by the National Science Foundation under grant CMMI-1826218 and the Air Force D3OM2S Center of Excellence under agreement FA8650-19-2-5209. 
A.J, C.C and S.P.O were supported by the Materials Project, funded by the U.S. Department of Energy, Office of Science, Office of Basic Energy Sciences, Materials Sciences and Engineering Division under contract no. DE-AC02-05-CH11231: Materials Project program KC23MP. 
S.J.L.B. was supported by the U.S. National Science Foundation through grant DMREF-1922234. A.A. and A.C. were supported by NIST award 70NANB19H005 and NSF award CMMI-2053929. 

\bibliography{sn-bibliography}% common bib file

\begin{thebibliography}{352}
\bibitem{nistdisclaimer}
\begin{botherref}
\oauthor{\bsnm{NIST\;disclaimer}}:
Certain commercial equipment, instruments, or materials are identified in this
  paper in order to specify the experimental procedure adequately. Such
  identification is not intended to imply recommendation or endorsement by
  NIST, nor is it intended to imply that the materials or equipment identified
  are necessarily the best available for the purpose.
\end{botherref}
\endbibitem
\end{thebibliography}


\end{document}